\documentclass[a4paper]{article}
\usepackage[a4paper,top=3cm,bottom=2cm,left=3cm,right=3cm,marginparwidth=1.75cm]{geometry}
\usepackage{amsfonts}
\usepackage{bm}
\usepackage{extarrows}
\usepackage{amssymb}
\usepackage{color}
\usepackage[all]{xy}
\usepackage{graphicx}
\usepackage{braket}
\usepackage{amsmath}
\usepackage{appendix}
\usepackage{graphicx}
\usepackage[colorinlistoftodos,prependcaption,textsize=tiny]{todonotes}
\usepackage{amsthm}
\usepackage{mathrsfs}
\usepackage{amssymb}
\usepackage{accents}
\usepackage{bbm}
\usepackage{extarrows}
\usepackage{makecell}
\usepackage{authblk}
\usepackage{url}
\usepackage{MnSymbol}
\usepackage{graphicx}
\usepackage{hyperref}
\usepackage{cite}
\usepackage{eufrak}
\hypersetup{colorlinks,
            citecolor=black,
            linkcolor=black,
            urlcolor=green,
            pdftex}

\newtheorem{theorem}{Theorem}

\def\be{\begin{equation}}
\def\ee{\end{equation}}
\def\ba{\begin{eqnarray}}
\def\ea{\end{eqnarray}}

\usepackage[english]{babel}
\usepackage[utf8x]{inputenc}
\usepackage[T1]{fontenc}

\usepackage[normalem]{ulem}

\usepackage{xargs}  
\newcommandx{\hgl}[2][1=]{\todo[linecolor=red,backgroundcolor=red!25,bordercolor=red,author=Hongguang, inline,#1]{#2}}

\title{Quantum representation of reduced twisted geometry in loop quantum gravity}
\author[1]{Gaoping Long }

\author[2]{Cong Zhang }
\author[3,4,5]{Hongguang Liu \footnote{corresponding author: liuhongguang@westlake.edu.cn}}

\affil[1]{College of Physics $\&$ Optoelectronic Engineering, Jinan University, Guangzhou, 510632, Guangdong, China}
\affil[2]{School of Physics and Astronomy, Key Laboratory of Multiscale Spin Physics,
Ministry of Education, Beijing Normal University, Beijing 100875, China}

\affil[3]{Institute for Theoretical Sciences, Westlake University, Hangzhou 310030, China
}

\affil[4]{Institute of Natural Sciences, Westlake Institute for Advanced Study, Hangzhou 310024, China
}

\affil[5]{Department Physik, Institut f\"ur Quantengravitation, Theoretische Physik III, Friedrich-Alexander Universit\"at Erlangen-N\"urnberg, Staudtstr. 7/B2, 91058 Erlangen, Germany}

\date{}

\begin{document}

\maketitle

\begin{abstract}
In this article, the quantum representation of the algebra among reduced twisted geometries (with respect to the Gauss constraint) is constructed in the gauge invariant Hilbert space of loop quantum gravity.  It is shown that the reduced twisted geometric variables not only describe the spatial discrete geometry more clearly, but also form a simple Poisson algebra which is analogous to that in quantum mechanics. By regularizing the reduced twisted geometric variables properly, the fundamental algebra of reduced twisted geometry is established, with the gauge invariant Hilbert space in loop quantum gravity as the corresponding quantum representation space.
This quantum representation also leads to fundamental operators associated with reduced twisted geometry. Based on these fundamental operators, a new type of extrinsic curvature operator is constructed in loop quantum gravity. 
\end{abstract}

\section{Introduction}
Loop quantum gravity (LQG) gives a convincing approach to the quantization of general relativity (GR) \cite{first30years,Ashtekar2012Background,RovelliBook2,Han2005FUNDAMENTAL,thiemann2007modern,rovelli2007quantum}. 
LQG is constructed by quantizing the connection dynamics formulation of the classical GR in loop representation. Specifically, the basic variables of the connection dynamics consist of the $SU(2)$ Ashtekar connection $A$ and its conjugate momentum $E$ representing the densitized triad. The loop quantum theory of this $SU(2)$ connection theory is constructed, by choosing the holonomies of $A$ along all edges and the fluxes of $E$ across all 2-surfaces as the fundamental variables. Crucially, the loop quantization is well-defined without requiring any extra background structures on the underlying spatial manifold. Hence, LQG inherits the background-independent nature of GR.
This theory has achieved several important breakthroughs (see, e.g. \cite{first30years} for a review and see, e.g., \cite{Zhang:2021qul,Long:2021izw,Zhang:2022vsl,Liegener:2019jhj,Han:2024rqb,Han:2024ydv,Assanioussi:2015gka} for some recent developments).
For instance, a family of operators representing spatial geometric observables (such as 2-surface area, 3-region volume, inverse metric tensor) has been regularized in LQG, without the need to subtract infinities. Moreover, the spectra of these spatial geometric operators have turned out to be discrete, predicting the fundamental discreteness of spatial geometry \cite{Ashtekar:1996eg,Ashtekar:1997fb,Bianchi:2008es,Ma:2010fy,Giesel_2006Consistencycheck,Yang_2019Consistencycheck,ROVELLI1995593,QoperatorPhysRevD.62.104021,volumePhysRevD.94.044003,long2020operators}. 
Furthermore, LQG offers a microscopic description of spacetime thermodynamics. The corresponding state counting, based on boundary states, and along with the computation of the entanglement entropy from quantum geometry states in LQG, successfully reproduce the black hole entropy, including specific quantum corrections \cite{Ashtekar:1997yu,Ashtekar:2000eq, Long:2024lbd,Song:2020arr,Kaul:2000kf,Ghosh:2011fc,Basu:2009cw,Engle:2010kt,Song:2022zit,Donnelly:2008vx,Perez:2014ura,Dasgupta:2005yu}. 
In addition, by applying the loop quantization method to symmetry-reduced models, such as in cosmology and black hole, LQG predicts intriguing phenomena like big bounce cosmology and black-to-white hole transitions \cite{Ashtekar:2003hd,Ashtekar:2011ni,Long:2020oma,Zhang:2021zfp,Bojowald:2001xe, Ashtekar:2006rx, Ashtekar:2006wn,Ashtekar:2005qt,Modesto:2005zm, Boehmer:2007ket,Chiou:2012pg,Gambini:2013hna,Corichi:2015xia,Dadhich:2015ora,Olmedo:2017lvt,Ashtekar:2018lag,BenAchour:2018khr,Han:2020uhb,Kelly:2020lec, Han:2022rsx,Giesel:2023hys,Ashtekar:2023cod,PhysRevLett.102.051301}.

The key feature of LQG is the discrete description of the space-time geometry. Specifically, the degrees of freedom of the discrete space-time geometry in LQG, including the spatial intrinsic geometry and extrinsic curvature on a spatial manifold, are encoded in the holonomy and flux variables. 
For a given graph upon which the holonomy and flux variables are defined,  the discrete geometry encoded in these holonomies and fluxes can be described by the so-called twisted geometry. The twisted geometry can be regarded as a generalized Regge geometry. More explicitly,  the spatial intrinsic geometry and extrinsic curvature in the twisted geometry are illustrated by the shapes and the gluing methods of the polyhedra which give a cellular partition of the spatial manifold.
Nevertheless, such a clear description of the discrete space-time geometry encoded in the holonomy and flux variables is only established in the classical level. If we turn to the quantum theory of LQG, this description is only half complete. More explicitly, 
the spatial intrinsic geometry and extrinsic curvature in quantum theory are governed by the corresponding geometric operators composed of the basic holonomy and flux operators. As mentioned above, the operators representing the spatial intrinsic geometric observables (2-surface area, 3-region volume, inverse metric tensor, etc.) have been studied thoroughly \cite{Ashtekar:1996eg,Ashtekar:1997fb,Bianchi:2008es,Ma:2010fy,Giesel_2006Consistencycheck,Yang_2019Consistencycheck,ROVELLI1995593,QoperatorPhysRevD.62.104021,volumePhysRevD.94.044003,long2020operators}.
However, the quantum description of spatial intrinsic geometric observables is not the full story of quantum geometry in LQG. In fact, the spatial extrinsic curvature operator is a key component in the scalar constraint operator, which governs the dynamics of LQG. In addition, extrinsic curvature is also included in the canonical expressions of ADM energy, expansion and surface gravity \cite{Yang:2008th,Nielsen:2007ac}, which are core topics in the study of black holes. Thus, the operator of extrinsic curvature is an essential ingredient in the study of LQG.
In fact, the study of the extrinsic curvature operator in LQG is rather inadequate. Though the extrinsic curvature operator can be established on the basis of the fundamental holonomy and flux operators, its specific expression is too complicated to analyze the properties of this operator. Nevertheless, the extrinsic curvature in the twisted geometry is equipped with a rather simple Poisson algebra, which inspires us to construct a new extrinsic curvature operator, to avoid the completeness of the previous one.  

 In this paper, we will establish the quantum representation of the Poisson algebra among the reduced twisted geometry variables, to give a full quantum description of the spatial intrinsic and extrinsic geometries. 
 Specifically,  we will first introduce the reduced twisted geometry variables to parametrize the gauge invariant twisted geometry, and  show that these reduced variables form a simpler Poisson algebra. Then, with the gauge invariant Hilbert space of LQG serving as  the quantum representation space, we will establish the fundamental algebra of reduced twisted geometry.  This quantum representation of the fundamental algebra also leads to the gauge invariant fundamental operators acting in the gauge invariant Hilbert space of LQG. Consequently, by expressing the extrinsic curvature in terms of the gauge invariant fundamental variables, we will construct a new type of extrinsic curvature operator accordingly.

This paper is organized as follows. In the following Sec. \ref{sec2}, we will review the kinematical structure and the existing treatment of extrinsic curvature in LQG. 
Then,  we will review the twisted geometry parameterization of the holonomy-flux phase space in Sec. \ref{sec:301}, and clarify the intrinsic geometry and extrinsic curvature encoded in the twisted geometry in Sec. \ref{sec:302};
Especially, in Sec. \ref{sec:303}, we will introduce the reduced twisted geometry, and show that the reduced twisted geometric variables form a Poisson algebra which is analogous to that in QM. Further, in Sec. \ref{sec:401} , the fundamental algebra of reduced twisted geometry will be constructed by requiring that its quantum representation generates the gauge invariant Hilbert space. Accordingly,  in Sec. \ref{sec:402},  the spatial extrinsic curvature operator will be constructed by expressing the extrinsic curvature in terms of the fundamental algebra of reduced twisted geometry.
  Finally, we will finish with a conclusion and discussion in Sec.\ref{sec5}.

\section{Elements of LQG}\label{sec2}
\subsection{The basic structures}
The (1+3)-dimensional Lorentzian LQG is constructed by canonically quantizing the classical GR, which is formulated as the connection dynamics based on the Yang-Mills phase space with the non-vanishing Poisson bracket\cite{Ashtekar:2004eh,Han2005FUNDAMENTAL}
 \begin{equation}
 \{A_{a}^i(x),E^{b}_j(y)\}=\kappa\beta\delta_a^b\delta^i_j\delta^{(3)}(x-y),
 \end{equation}
where the configuration and momentum variables are respectively the  $su(2)$-valued connection field $A_{a}^i$ and densitized triad field $E^{b}_j$ on a 3-dimensional spatial manifold $\Sigma$, $\kappa$ and $\beta$ represent the gravitational constant and Babero-Immirze parameter respectively. Here we use $i, j, k, ...$ for the internal $su(2)$ index and $a, b, c, ...$ for the spatial index. Let $q_{ab}=e_a^ie_{bi}$ be the spatial metric on $\Sigma$. The densitized triad $E^{b}_j$ is related to the triad $e_i^a$ by $E^{a}_{i}=\sqrt{\det(q)}e^{a}_{i}$, where $\det(q)$ denotes the determinant of $q_{ab}$. The connection can be decomposed as $A_{a}^{i}=\Gamma_{a}^{i}+\beta K_{a}^{i}$, where $\Gamma_{a}^{i}$ is the Levi-Civita connection of $e_{a}^{i}$, which is given by \cite{thiemann2008modern}
\begin{equation}
\Gamma_{a}^{i}=\frac{1}{2}\epsilon^{ijk}e_k^b
(\partial_be_a^j-\partial_ae_b^j+e^c_je_{al}\partial_be^l_c);
\end{equation}
 $K_a^i$ is related to the extrinsic curvature $K_{ab}$ by $K_a^i=K_{ab}e^b_j\delta^{ji}$. The dynamics of GR in this Yang-Mills phase space is governed by the following Gaussian, diffeomorphism and scalar constraints, which are given by
\begin{equation}\label{GC}
 \mathcal{G}:=\partial_aE^{ai}+A_{aj}E^a_k\epsilon^{ijk}\approx0,
\end{equation}
\begin{equation}\label{VC}
 \mathcal{C}_a:=E^b_iF^i_{ab}\approx0,
\end{equation}
and
\begin{eqnarray}\label{SC}
\mathcal{C}:=\frac{E^a_i E^b_j}{\sqrt{|\det{(E)}|}}({\epsilon^{ij}}_kF^k_{ab}-2(1+\beta^2)K^i_{[a}K^j_{b]})\approx0
\end{eqnarray}
respectively, where $F_{ab}^i=\partial_aA_b^i-\partial_bA_a^i+\epsilon_{ijk}A_a^jA_b^k$ is the curvature of $A_a^i$. Equivalently, the scalar constraint is also given by \cite{Alesci:2015wla,Assanioussi:2015gka}
\begin{equation}\label{SC2}
\mathcal{C}:=-\frac{1}{\beta^2}\frac{E^a_i E^b_j}{\sqrt{|\det{(E)}|}}{\epsilon^{ij}}_kF^k_{ab}-(1+\frac{1}{\beta^2})\sqrt{|\det{(E)}|}R\approx0,
\end{equation}
where  $\sqrt{|\det{(E)}|}R:=-\sqrt{|\det{(E)}|}R_{ab}^j\epsilon_{jkl}e^{ak}e^{bl}$ is the  densitized scalar curvature of the spatial metric $q_{ab}$, with  $R_{ab}^j:=2\partial_{[a}\Gamma_{b]}^j+\epsilon^j_{\ kl}\Gamma_{a}^k\Gamma^l_b$ \cite{thiemann2008modern}.

In the Yang-Mills phase space coordinated by $(A_{a}^i,E^{b}_j)$, the gauge invariant (with respect to Gaussian constraint) degrees of freedom are completely contained in the geometric variables $q_{ab}$ and $K_{cd}$.  Notice that the spatial metric $q_{ab}$ can be given by the densitized triad field $E^{b}_j$ directly, i.e. $\det(q)q^{ab}=E^a_iE^{bi}$. However, it is not intuitive to express the extrinsic curvature $K_{ab}=K_{ai}e_{b}^i$ in terms of the  basic variables $A_{a}^i$ and $E^{b}_j$.  Especially,  one can also notice that the extrinsic curvature one-form $K_a^i$ appears in the expression \eqref{SC}
 of scalar constraint. Hence, it is necessary to express $K_a^i$ in terms of the basic variables $A_{a}^i$ and $E^{b}_j$.  In the existing literature, the extrinsic curvature one-form $K_a^i$ can be expressed in terms of the basic variables $A_{a}^i$ and $E^{b}_j$ as
 \begin{eqnarray}\label{K1form}
 K_a^i=\frac{1}{\kappa\beta}\{A_a^i,K\}
\end{eqnarray}
where 
 
  \begin{eqnarray}
 K:=\int _{\Sigma}dx^3 K_a^iE^a_i=\{\int d^3x\frac{\mathcal{C}_{\text{E}}(x)}{2\kappa}, V(\Sigma)\},
\end{eqnarray}
with $\mathcal{C}_{\text{E}}(x):=\frac{E^a_i E^b_j}{\sqrt{|\det{(E)}|}}{\epsilon^{ij}}_kF^k_{ab}$, and  the volume $V(R)$ of an open region $R$ in $\Sigma$ being defined by $V(R):=\int_{R}d^3x\sqrt{|\det(E)|}$. 

The loop quantization of the $SU(2)$ connection formulation of GR leads to a kinematical Hilbert space $\mathcal{H}$, which can be regarded as a union of the Hilbert spaces $\mathcal{H}_\gamma=L^2((SU(2))^{|E(\gamma)|},d\mu_{\text{Haar}}^{|E(\gamma)|})$ on all possible finite graphs $\gamma$,  where $E(\gamma)$ denotes the set of independent edges in $\gamma$,  $|E(\gamma)|$ denotes the number of elements in $E(\gamma)$, and $d\mu_{\text{Haar}}^{|E(\gamma)|}$ denotes the product of the Haar measure on $SU(2)$. In this sense, on each given $\gamma$ there is a discrete phase space $(T^\ast SU(2))^{|E(\gamma)|}$, which is coordinatized by the basic discrete variables---holonomies and fluxes. The holonomy of $A_a^i$ along an edge $e\in E(\gamma)$ is defined by
 \begin{equation}
h_e[A]:=\mathcal{P}\exp(\int_eA)=1+\sum_{n=1}^{\infty}\int_{0}^1dt_n\int_0^{t_n}dt_{n-1}...\int_0^{t_2} dt_1A(t_1)...A(t_n),
 \end{equation}
 where $A(t)=A_a^i(t)\dot{e}^a(t)\tau_i$, and $\tau_i=-\frac{\textbf{i}}{2}\sigma_i$ with $\sigma_i$ being the Pauli matrices. The gauge covariant flux  $F^i_e$  of $E^b_j$  through the 2-face dual to edge $e\in E(\gamma)$ is defined by  \cite{Zhang:2021qul,Thomas2001Gauge}
 \begin{equation}\label{F111}
 F^i_e:=-\frac{2}{\beta }\text{tr}\left(\tau^i\int_{S_e}\epsilon_{abc}h(\rho_{s(e)}(\sigma))E^{cj}(\sigma)\tau_jh(\rho_{s(e)}(\sigma)^{-1})\right)
 \end{equation}
 in the perspective of source point of $e$,
 where $S_e$ dual to $e$ is the 2-face in the dual lattice $\gamma^\ast$ of $\gamma$, $\rho_{s(e)}(\sigma): [0,1]\rightarrow \Sigma$ is a path connecting the source point $s(e)\in e$ to $\sigma\in S_e$ such that $\rho_{s(e)}(\sigma): [0,\frac{1}{2}]\rightarrow e$ and $\rho_{s(e)}(\sigma): [\frac{1}{2}, 1]\rightarrow S_e$.  
 Similarly, the corresponding flux $\tilde{F}^i_e$ in the perspective of target point of $e$ is defined by
 \begin{equation}\label{F222}
 \tilde{F}^i_e:=\frac{2}{\beta }\text{tr}\left(\tau^i\int_{S_e}\epsilon_{abc}h(\rho_{t(e)}(\sigma))E^{cj}(\sigma)\tau_jh(\rho_{t(e)}(\sigma)^{-1})\right),
 \end{equation}
 where $\rho_{t(e)}(\sigma): [0,1]\rightarrow \Sigma$ is a path connecting the target point ${t(e)}\in e$ to $\sigma\in S_e$ such that $\rho_{t(e)}(\sigma): [0,\frac{1}{2}]\rightarrow e$ and $\rho_{t(e)}(\sigma): [\frac{1}{2}, 1]\rightarrow S_e$. It is easy to see that one has the relation
 \begin{equation}\label{F333}
 \tilde{F}^i_e\tau_i=-h_e^{-1} {F}^i_e\tau_ih_e.
 \end{equation}
 The non-vanishing Poisson brackets among the holonomy and fluxes read
 \begin{eqnarray}\label{hp1220}
&&\{{{h}}_e[{A}],{{F}}^i_{e'}\}= \delta_{e,e'}{\kappa}\tau^i{{h}}_e[{A}],
\quad \{{{F}}^i_{e},{{F}}^j_{e'}\}= \delta_{e,e'}{\kappa}{\epsilon^{ij}}_k{{F}}^k_{e'},\\\nonumber
&&\{{{h}}_e[{A}],\tilde{{F}}^i_{e'}\}=- \delta_{e,e'}{\kappa}{{h}}_e[{A}]\tau^i,
\quad \{\tilde{{F}}^i_{e},\tilde{{F}}^j_{e'}\}=- \delta_{e,e'}{\kappa}{\epsilon^{ij}}_k{\tilde{F}}^k_{e'}.
\end{eqnarray}

The basic operators in $\mathcal{H}_\gamma$ are given by promoting the basic discrete variables as operators. The resulting holonomy and flux operators act on cylindrical functions $f_\gamma(A)=f_\gamma(h_{e_1}[A],...,h_{e_{|E(\gamma)|}}[A])$ in $\mathcal{H}_\gamma$ as
\begin{equation}
\hat{h}_e[A]f_\gamma(A)=h_e[A]f_\gamma(A),
\end{equation}
\begin{equation}
\hat{F}^i_ef_\gamma(h_{e_1}[A],...,h_{e}[A],...,h_{e_{|E(\gamma)|}}[A])=-\textbf{i}\kappa\hbar\frac{d}{d\lambda} f_\gamma\left.\left(h_{e_1}[A],...,e^{\lambda\tau^i}h_{e}[A],...,h_{e_{|E(\gamma)|}}[A]\right)\right|_{\lambda=0},
\end{equation}
\begin{equation}
\hat{\tilde{F}}^i_ef_\gamma(h_{e_1}[A],...,h_{e}[A],...,h_{e_{|E(\gamma)|}}[A])=\textbf{i}\kappa\hbar\frac{d}{d\lambda} f_\gamma\left.\left(h_{e_1}[A],...,h_{e}[A]e^{\lambda\tau^i},...,h_{e_{|E(\gamma)|}}[A]\right)\right|_{\lambda=0}.
\end{equation}
Two spatial geometric operators in $\mathcal{H}_\gamma$ are worth to mention here. The first one is the area operator defined as $|\beta \hat{F}_e|:=\sqrt{\beta^2\hat{F}^i(e)\hat{F}_i(e)}$ , which represents the area of the 2-face $S_e$. As a remarkable prediction of LQG, the area operator take the following discrete spectrum \cite{Han2005FUNDAMENTAL},
 \begin{equation}\label{eigenp0}
 \text{Spec}(|\beta \hat{F}_e|)=\{\beta \kappa\hbar\sqrt{j(j+1)}|j\in\frac{\mathbb{N}}{2}\}.
 \end{equation}
 The second important spatial geometric operator is the volume operator of an open region $R\subset \Sigma$. The volume ${V}_R$ of the open region $R$ is $$V_R=\int_{R} dx^3 \sqrt{|\det(q)|},$$ where $\det(q)$ is the determinant of the spatial metric $q_{ab}$. Then, the corresponding operator of  $V_R$ is defined as \cite{Ashtekar:1997fb}
\begin{equation}\label{Vdef}
\hat{V}_R f_\gamma:=\sum_{v\in V(\gamma)\cap R}\hat{V}_vf_\gamma=\sum_{v\in V(\gamma)\cap R}\sqrt{|\hat{Q}_v|}f_\gamma,
\end{equation}
where $V(\gamma)$ denotes the set of vertices of $\gamma$, and
\begin{equation}
\hat{Q}_v:=\frac{1}{8}(\beta )^3\sum_{\{e_I,e_J,e_K\}\subset E(\gamma)}^{e_I\cap e_J\cap e_K=v}\epsilon_{ijk}\epsilon^{IJK}\hat{F}^i(v,e_I)\hat{F}^j(v,e_J)\hat{F}^k(v,e_K),
\end{equation}
where $\epsilon^{IJK}=\text{sgn}[\det(e_I\wedge e_J\wedge e_K)]$, $\hat{F}^i(v,e)=\hat{F}^i(e)$ if $s(e)=v$ and $\hat{F}^i(v,e)=-\hat{\tilde{F}}^i(e)$ if $t(e)=v$.

As we mentioned above, the quantization of several valuable observables in GR involves the operator corresponding to the extrinsic curvature 1-form $K_a^i$. In fact, there is already a strategy to construct  such an operator. Specifically, one has the operator corresponding to the smearing variable $$K_e=\int_{e}dtK_a^i(t)\dot{e}^a(t)\tau_i,$$ which reads
\begin{equation}\label{Kexist}
\hat{ K}_e=-\frac{1}{2\beta\kappa^2\hbar^2}\sum_{v\in V(\gamma)}\sum^{e_I\cap e_J\cap e_K=v}_{\{e_I,e_J,e_K\}\subset E(\gamma)}\epsilon^{IJK}\text{tr}\left(h_{e_I}[h^{-1}_{e_I},[\hat{C}_E[1],\hat{V}_v]] \right),
\end{equation}
where $\hat{C}_E[1]$ is given by the Euclidean part of the scalar constraint operator; For the special model of non-graph-changing and cubic graph $\gamma$, this Euclidean part is defined as
\begin{equation}
\hat{\mathcal{C}}_E{[N]}=-\frac{4}{\textbf{i}\beta \kappa\hbar}\sum_{v\in V(\gamma)}N(v)\sum^{e_I\cap e_J\cap e_K=v}_{\{e_I,e_J,e_K\}\subset E(\gamma)}\epsilon^{IJK}\text{tr}(h_{\alpha_{IJ}}h_{e_K}[h^{-1}_{e_K},\hat{V}_v]),
\end{equation}
where $e_I,e_J, e_K$ have been re-oriented to be outgoing at $v$, 
$\alpha_{IJ}$ is the minimal loop around a plaquette containing $e_I$ and $e_J$ \cite{Han:2020chr,Giesel_2007}, which begins at $v$ via $e_I$ and gets back to $v$ through $e_J$. 

The operator $\hat{ K}_e$ is well-defined in the Hilbert space of LQG. However, the action of $\hat{ K}_e$ on the quantum state is complicated since it is not a polynomial of the basic holonomy and flux operators, so that it is hard to carry out the explicit calculation. Especially, this  also present  an obstacle to  the constructions and calculations of the operators which contain the extrinsic curvature, e.g, scalar constraint operators, ADM energy operator and expansion operator. In the following part of this paper,  we will consider the quantum representation of the reduced twisted geometry, upon which the extrinsic curvature can be regularized and quantized in a new strategy.

\section{Twisted geometry and its reduction}

\subsection{Twisted geometric parametrization of $SU(2)$ holonomy-flux phase space}\label{sec:301}
The holonomy-flux phase space associated to a given graph $\gamma$ is coordinatized by the classical holonomies and fluxes. The discrete geometry information of the dual lattice of $\gamma$  is encoded in the holonomy-flux variables, which can be explained by the so-called twisted geometry parametrization of the holonomy-flux phase space \cite{PhysRevD.82.084040,PhysRevD.103.086016}.
Let us give a brief introduction of this parametrization as follows.

From now on, let us focus on a graph $\gamma$ whose dual lattice gives a partition of $\Sigma$ constituted by 3-dimensional polytopes. Elementary edge are referred to as those passing through exactly one 2-dimensional face in the dual lattice of $\gamma$. The discrete phase space related to the given graph $\gamma$ is  $\times_{e\in E(\gamma)}T^\ast SU(2)_e$ with $e$ being the elementary edges of $\gamma$. The symplectic potential on the phase space reads
\begin{equation}\label{sym1}
  \Theta_{\gamma}=\frac{a^2}{\kappa}\sum_{e\in E(\gamma)}\text{Tr}(p_e^i\tau_idh_eh_e^{-1}),
\end{equation}
where $p_e^i:=\frac{F_e^i}{a^2}$ is the dimensionless
flux with $a$ being a constant with the dimension of length, and $\text{Tr}(XY):=-2\text{tr}_{1/2}(XY)$ with $X,Y\in su(2)$.
Without loss of generality, we can first focus on the space $T^\ast SU(2)_e$ associated to one single elementary edge $e\in E(\gamma)$.
This space can be parametrized by the so-called twisted geometry variables
 \begin{equation}
 (V_e,\tilde{V}_e,\xi_e, \eta_e)\in P_e:=S^2_e\times S^2_e\times T^\ast S^1_e,
 \end{equation}
 where $\eta_e\in\mathbb{R}$,  $\xi_e\in [0,2\pi)$, and
  \begin{equation}
  V_e:=V_e^i\tau_i, \   \ \  \tilde{V}_e:=\tilde{V}^i_e\tau_i,
  \end{equation}
 with $V_e^i,\tilde{V}^i_e\in S^2_e$ being unit vectors. 
 To relate the holonomy-flux variables with the twisted geometry variables, we need to specify two sections $n_e:S^2_e\to SU(2)$ and $\tilde{n}_e:S^2_e\to SU(2)$ such that $V^i_e\tau_i=n_e(V_e)\tau_3n_e(V_e)^{-1}$ and $\tilde{V}^i_e\tau_i=-\tilde{n}_e(\tilde{V}_e)\tau_3\tilde{n}_e(\tilde{V}_e)^{-1}$. Then, the twist geometry variables can be related to the holonomy-flux by the map
\begin{eqnarray}\label{para}
(V_e,\tilde{V}_e,\xi_e,\eta_e)\mapsto(h_e,p^i_e)\in T^\ast SU(2)_e:&& p^i_e\tau_i=\eta_e V_e=\eta_en_e(V_e)\tau_3n_e(V_e)^{-1},\\\nonumber
&&h_e=n_e(V_e)e^{\xi_e\tau_3}\tilde{n}_e(\tilde{V}_e)^{-1}.
\end{eqnarray}
Note that this map is a two-to-one map. In other words, \eqref{para} will map two points $\pm (V_e,\tilde{V}_e,\xi_e,\eta_e)\in P_e$ to the same point $(h_e,p^i_e)\in T^\ast SU(2)_e$.
Hence, by selecting either branch between the two signs related by a $\mathbb{Z}_2$ symmetry, one can establish a bijection between $\left.T^*SU(2)_e\right|_{|p_e^i|>0}$ and the region in $P_e$ with either $\eta_e>0$ or $\eta_e<0$ \cite{PhysRevD.82.084040,PhysRevD.103.086016}.

 Now we can return to the discrete phase space of LQG on the whole graph $\gamma$. It is just the Cartesian product of the discrete phase space on each single elementary edge of $\gamma$. That is, the discrete phase space on $\gamma$ can be re-parametrized by using the twisted geometry variables in $P_\gamma:=\times_{e\in E(\gamma)}P_e$. To see how the variables in $P_\gamma$ describe the geometry associated to $\gamma$, we note that $|\eta_e|$ is interpreted as the dimensionless area of the 2-dimensional face dual to $e$, leading to the interpretation of $\eta _e V_e$ and $\eta _e \tilde{V}_e$ as the area-weighted outward normal vectors of the face in the frames of source and target points of $e$ respectively.
 Then, the holonomy $h_e=n_e(V_e)e^{\xi_e\tau_3}\tilde{n}^{-1}_e(\tilde{V}_e)$ rotates the inward normal $-\eta _e\tilde{V}_e$ to the outward normal $\eta _e{V}_e$ as
  \begin{equation}
  \tilde{V}_e=-h_e^{-1}V_eh_e.
  \end{equation} 
  The symplectic 1-form on $P_\gamma$ resulting from \eqref{sym1} and the re-parametrization \eqref{para} reads
  \begin{equation}\label{sym2}
  \Theta_{P_\gamma}=\frac{a^2}{\kappa}\sum_{e\in E(\gamma)}\left(\eta_e\text{Tr}(V_edn_en_e^{-1})+\eta_ed\xi_e +\eta_e\text{Tr}(\tilde{V}_ed\tilde{n}_e\tilde{n}_e^{-1})\right),
\end{equation}
where $\text{Tr}(V_edn_en_e^{-1})$ and $\text{Tr}(\tilde{V}_ed\tilde{n}_e\tilde{n}_e^{-1})$ are the standard symplectic 1-form on the unit sphere $S^2_e$ \cite{PhysRevD.82.084040,PhysRevD.103.086016}.  In the following part of this article, we will focus on the branch $P_\gamma^+:=\times_{e\in E(\gamma)}P_e^+$,  with $P_e^+:=(S^2_e\times S^2_e\times \mathbb{R}^+_{e}\times S^1_e)$ being the branch of $P_e$ satisfying $\eta_e\geq0$.

It is necessary to consider the gauge reduction with respect to the Gauss constraint in the twisted geometry space $P_\gamma^+$. The discrete Gauss constraint in the holonomy-flux phase
space $\times_{e\in E(\gamma)}T^\ast SU(2)_e$ is given by
 \begin{equation}
G_v:= -\sum_{e,s(e)=v}{p}^i_e\tau_i+\sum_{e,t(e)=v}p^i_eh_e^{-1}\tau_ih_e=0.
 \end{equation}
 The gauge transformation $\{g_v|v\in V(\gamma)\}$ at the vertices of $\gamma$ generated by the Gauss constraint is given by
 \begin{equation}
 h_e\mapsto g_{s(e)}h_eg_{t(e)}^{-1},\quad {p}^i_e\tau_i\mapsto {p}^i_e g_{s(e)}\tau_i g_{s(e)}^{-1}.
 \end{equation}
 Correspondingly, this gauge transformation can be expressed in terms of the twisted geometry variables by using the re-parametrization \eqref{para} , which reads
 \begin{eqnarray}\label{gaugetransV}
  {V}^i_e\tau_i\mapsto V_e(g_{s(e)}):= {V}^i_e g_{s(e)}\tau_i g_{s(e)}^{-1},  &&  \tilde{V}^i_e\tau_i\mapsto \tilde{V}_e(g_{t(e)}):=\tilde{V}^i_e g_{t(e)}\tau_i g_{t(e)}^{-1},\\\nonumber
  \xi_e\mapsto   \xi_e+\xi_{g_{s(e)}}-\xi_{g_{t(e)}}, &&  \eta_e\mapsto\eta_e,
 \end{eqnarray}
 where $\xi_{g_{s(e)}}$ and $\xi_{g_{t(e)}}$ are determined respectively by
 \begin{eqnarray}\label{xige}
 && g_{s(e)}n_e(V_e)=n_e(V_e(g_{s(e)}))e^{\xi_{g_{s(e)}}\tau_3},\\\nonumber
  &&g_{t(e)}\tilde{n}_e(\tilde{V}_e)=\tilde{n}_e(\tilde{V}_e(g_{t(e)}))e^{\xi_{g_{t(e)}}\tau_3}.
 \end{eqnarray}
Then, the gauge reduction with respect to the discrete Gauss constraint can be carried out in $P_\gamma^+:=\times_{e\in E(\gamma)}P_e^+$, and the resulting reduced phase space is given by
 \begin{equation}
{H}^+_\gamma:={P}^+_\gamma/\!/SU(2)^{|V(\gamma)|}=\left(\times_{e\in E(\gamma)} T^\ast S_e^1\right)^+\times \left(\times_{v\in V(\gamma)} \mathfrak{P}_{\vec{\eta}_v}\right)
\end{equation}
  where $V(\gamma)$ is the collection of vertices in $\gamma$,  $|V(\gamma)|$ is the number of the vertices in $\gamma$ and
 \begin{equation}
 \mathfrak{P}_{\vec{\eta}_v}:=\{(V_{e_1},...,V_{e_{m_v}},\tilde{V}_{e_1},...,\tilde{V}_{e_{\tilde{m}_v}})\in \times_{e| s(e)=v}S^{2}_{e} \times_{e| t(e)=v}S^{2}_{e}| G_{v}=0 \}/SU(2);
 \end{equation}
Here $s(e)$ and $t(e)$ represent the source point and the target point of $e$ respectively, $m_v$ is the number of the edges starting at $v$, and $\tilde{m}_v$ is the number of the edges ending at $v$.

\subsection{Intrinsic geometry and extrinsic curvature in twisted geometry}\label{sec:302}

Similar to the connection phase space, the reduced twisted geometry contains complete discrete spatial geometry associated to $\gamma$, i.e., includes the spatial intrinsic and extrinsic geometry. The spatial intrinsic geometries associated to $\gamma$ are given by various discrete geometry variables, e.g. area, volume, length, angle and the polyhedra geometry on each vertex \cite{Ashtekar:1996eg,Ashtekar:1997fb,Bianchi:2008es,Ma:2010fy,Giesel_2006Consistencycheck,Yang_2019Consistencycheck,ROVELLI1995593,QoperatorPhysRevD.62.104021,volumePhysRevD.94.044003,long2020operators,PhysRevD.83.044035,Long:2020agv}.  All of these discrete spatial geometry variables are constructed by  the fluxes purely, and thus, they can be re-expressed in terms of the reduced twisted geometry variables by using the relations $p^i_e=\eta_e V^i_e$ and $\tilde{p}^i_e=\eta_e \tilde{V}^i_e$ .
However, this is not the full story of the reduced twisted geometry, since the extrinsic curvature of the discrete spatial geometry is still not clear in the framework of the twisted geometry. In fact,   to extract the information of extrinsic curvature from the twisted geometry, it is necessary to give a clear expression of the holonomy of spin connection \cite{PhysRevD.87.024038,Long:2024pxb}.

Now, let us start to introduce the holonomy of spin connection in twisted geometry  by separating the degrees of freedom of the intrinsic and extrinsic geometry encoded in the holonomy $h_e$ of Ashtekar connection. First, let us focus on the Regge geometry which is a sub-sector of twisted geometry.
As shown in Ref.\cite{Rovelli:2010km}, the discrete extrinsic curvature in the Regge geometry  is given by a distribution field
\begin{equation}
k_{ab}(x)=k_{ab}\int_{S_e}\delta^3(x,S_e(\sigma))d^2\sigma,
\end{equation}
where $S_e$ is a 2-dimensional face shared by two adjacent polyhedra, $\{x\}$ is a coordinate system cover the face $S_e$, $\{\sigma\}$ is a coordinate system on $S_e$, and
\begin{equation}
k_{ab}(x)=\alpha_e V_aV_b
\end{equation}
with $V_a$ being the 3-D unit normal co-vector of the face $S_e$, and $\alpha_e$ being the dihedral
angle between the 4-D normals of the two adjacent polyhedra at $S_e$. Notice that the two adjacent polyhedra in the Regge geometry are aligned at $S_e$. Thus,  one can always find a local cartesian coordinate system covering these two adjacent polyhedra, in which $e^a_i=\delta^a_i$ is satisfied so that the spin connection vanishes.  By considering an edge $e$ which crosses and only crosses the face $S_e$,  the Ashtekar holonomy over $e$ can be given by
\begin{equation}
h_e=\mathcal{P}\exp(\beta \int_e dt \delta^{bi}k_{ab}(t)\dot{e}^a(t)\tau_i)=\exp(\beta \alpha_e V_e^i\tau_i),
\end{equation}
where $V_e^i=V_a e^{ai}$. So far we have worked in the Regge geometry, in which  a gauge is fixed to ensure that the two polyhedra adjacent at the face $S_e$ are also aligned at $S_e$ with a matched shape. Then,  let us extend the consideration to the twisted geometry, by rotating the polyhedron containing the target point $t(e)$ of $e$ with an arbitrary $SU(2)$ rotation $h^{\Gamma}_e$, and releasing the shape matching condition at $S_e$ \cite{Rovelli:2010km}. Now, the Ashtekar holonomy  gets
an additional contribution, which reads
\begin{equation}\label{he222}
h_e=\exp(\beta \alpha_e V_e^i\tau_i)h^{\Gamma}_e=n_e(V_e)e^{(\zeta_e+\beta\alpha_e) \tau_3}\tilde{n}_e(\tilde{V}_e)^{-1},
\end{equation}
where we used the decomposition 
\begin{equation}\label{hnn}
h_e^{\Gamma}=\exp(\zeta_eV_e^i\tau_i)n_e(V_e)\tilde{n}_e(\tilde{V}_e)^{-1}=n_e(V_e)e^{\zeta_e \tau_3}\tilde{n}_e(\tilde{V}_e)^{-1}
\end{equation}
with $  V_e=V_e^i\tau_i,\tilde{V}_e=\tilde{V}^i_e\tau_i$.
In fact,  $h_e^{\Gamma}$ introduced here is just the holonomy of spin connection in the twisted geometry. Also,  by comparing Eq.\eqref{para} and Eq.\eqref{he222}, one has 
 \begin{eqnarray}
\label{decomp3}
\xi_e=\zeta_e+\beta\alpha_e.
\end{eqnarray}
Moreover,  by following the gauge transformation of $\{(\xi_e,V_e, \tilde{V}_e)|e\in E(\gamma)\}$ given by Eqs.\eqref{gaugetransV}, it is easy to see that the holonomy $h^{\Gamma}$ of spin connection is transformed as
 \begin{equation}\label{gaugetranshgamma}
 h^{\Gamma}_e\mapsto g_{s(e)}h^{\Gamma}_eg_{t(e)}^{-1};
 \end{equation}
Correspondingly, $\zeta_e$ is transformed as
\begin{equation}\label{zetatrans}
\zeta_e\mapsto   \zeta_e+\xi_{g_{s(e)}}-\xi_{g_{t(e)}}
\end{equation}
with $\xi_{g_{s(e)}}$ and $\xi_{g_{t(e)}}$ being given by Eq.\eqref{xige}.

Notice that $n_e(V_e)$ and $\tilde{n}_e(\tilde{V}_e)$ in the expression  \eqref{hnn} of $h_e^{\Gamma}$ are determined by the fluxes clearly. Also, it is worth to giving the specific expression of $\zeta_e$ in $h_e^{\Gamma}$. 
Let us focus on the cubic graph $\gamma$ on the manifold with topology $\mathbb{T}^3$. Since the gauge transformation of $\zeta_e$ is known, we just need to define $\zeta_e$ in a specific gauge.  Let us choose the gauge at $s(e)$ and $t(e)$ which ensures $V_e=-\tilde{V}_e$;  Now,  $e^{\zeta_e V_e}$ is just the $SU(2)$ element which rotates the polyhedron containing $t(e)$ to ensure it aligns with the polyhedron containing $s(e)$ at their faces dual to $e$. However, an issue arises at this stage for the twisted geometry. Notice that the faces of these two adjacent polyhedra dual to $e$ may have un-matched shapes, which leads that the alignment of these two adjacent polyhedra at their glued faces dual to $e$ becomes ambiguous. To avoid this issue, one can choose a relaxed alignment strategy adapted to arbitrary polyhedrons, in which only  one pair of edges in the glued faces are aligned.  Notice that the aligned edges in the glued faces are dual to  a minimal loop $\square_e$ in $\gamma$ containing $e$, thus the expression of $\zeta_{e}$ in the holonomy of spin connection $h^{\Gamma}_{e}= n_e e^{\zeta_{e}\tau_3}\tilde{n}_e^{-1}$
depends on the choice of the minimal loop $\square_e$.
Now, by choosing  a minimal loop $\square_e$ containing $e$, $e_s$ and $e_t$ with $s(e_s)=s(e)$ and $s(e_t)=t(e)$, the angle $\zeta_e=\zeta_{e,\square_e}\in(-\pi,\pi]$ can be determined by 
\begin{equation}
\cos(\zeta_e)=\frac{\delta^{ii'}\epsilon_{ijk}V^j_{e_s}V^k_e\epsilon_{i'j'k'}\tilde{V}^{j'}_{e}V^{k'}_{e_t}}{|\epsilon_{ijk}V^j_{e_s}V^k_e|\cdot|\epsilon_{i'j'k'}\tilde{V}^{j'}_{e}V^{k'}_{e_t}|}
\end{equation}
and 
\begin{equation}
\text{sgn}(\zeta_e)=\text{sgn}(\epsilon_{ijk}V_{e_s}^iV_{e_t}^jV_e^k)
\end{equation}
for the gauge aforementioned
which ensures $V_e=-\tilde{V}_e$. In the following part of this article, we will focus on the the cubic graph $\gamma$ on the manifold with topology $\mathbb{T}^3$.

It is ready to turn to consider the extrinsic curvature in the twisted geometry.  Recall the expressions \label{he222} of the Ashtekar holonomy $h_e$ and the decomposition \label{hnn} of the holonomy $h_e^\Gamma$ of spin connection. The contribution of extrinsic curvature to $h_e$  can be expressed in either the local gauge for the source polyhedra or that for the target polyhedra, respectively as  \cite{Rovelli:2010km}
\begin{eqnarray}
\label{corres2}
e^{\beta K_{e,s(e)}}:=e^{\varrho_e {V}_e}\equiv e^{\beta\alpha_e {V}_e}=h_e(h^\Gamma_e)^{-1} \,\,\,\text{or}\,\, \,e^{\beta K_{e,t(e)}}:=e^{-\varrho_e \tilde{V}_e}\equiv e^{-\beta \alpha_e \tilde{V}_e}=(h^\Gamma_e)^{-1}h_e,
\end{eqnarray}
where we used the notation 
\begin{equation}
\varrho_e\equiv\beta \alpha_e=\xi_e- \zeta_e.
\end{equation}
 Then, one has the following correspondence 
\begin{eqnarray}
\label{corres3}
 K_{e ,s(e)}=\frac{1}{\beta}\varrho_e {V}_e\,\,\,\text{or}\,\, \, K_{e ,t(e)}=-\frac{1}{\beta}\varrho_e\tilde{V}_e
\end{eqnarray}
 in the perspectives of source frame or target frame of $e$ respectively.
Now, the above procedures for the extrinsic data extracted from twisted geometry suggest us a new regularization method of the extrinsic curvature on a graph $\gamma$.  Let us define 
\begin{equation}\label{regu1form}
 \mathcal{K}^i_{e,s(e)}:=\mathcal{K}_eV_e^i \ \text{and}\  \mathcal{K}^i_{e,t(e)}:=-\mathcal{K}_e\tilde{V}_e^i
\end{equation}
with
\begin{equation}
\mathcal{K}_e:=\frac{(e^{\textbf{i}r(\eta_e)\varrho_e}-e^{-\textbf{i}r(\eta_e)\varrho_e})}{2\textbf{i}r(\eta_e)\beta} ,
\end{equation}
where $r(\eta_e):=\frac{\sqrt{({\eta}_{e})^2+t^2/4}}{\eta_e}$ is a regulator and we will show its property in next section. By  assuming $|\varrho_e|\propto \mu$ with $\mu$ being the coordinated length of $e$, it is easy to check that
\begin{equation}
\dot{e}^a K_a^i\tau_i|_{p(e)}=\lim_{\mu\to 0}\frac{K_{e ,s(e)}}{\mu}=\lim_{\mu\to 0}\frac{\mathcal{K}^i_{e,s(e)}\tau_i}{\mu}
\end{equation}
or
\begin{equation}
\dot{e}^a K_a^i\tau_i|_{p(e)}=\lim_{\mu\to 0}\frac{K_{e ,t(e)}}{\mu}=\lim_{\mu\to 0}\frac{\mathcal{K}^i_{e,t(e)}\tau_i}{\mu},
\end{equation}
where we used that \cite{Long:2024pxb}
\begin{equation}
\lim_{\mu\to 0} e=p(e),\quad  \lim_{\mu\to 0}\frac{h_e-\mathbb{I}}{\mu}=\dot{e}^aA_a^i\tau_i|_{p(e)},\quad   \lim_{\mu\to 0}\frac{h^\Gamma_e-\mathbb{I}}{\mu}=\dot{e}^a\Gamma_a^i\tau_i|_{p(e)}.
\end{equation}
Thus,  $\mathcal{K}^i_{e,s(e)}$ or $\mathcal{K}^i_{e,t(e)}$ can be regarded as a new regularized version of $K_a^i$ along the edge $e$, when it is expressed in the perspectives of source frame or target frame of $e$ respectively.
Based on these conventions, the densitized extrinsic curvatures on the graph $\gamma$ can be defined by
\begin{eqnarray}\label{Kdef1}
 \mathcal{K}_e^{\ e}&:=&\mathcal{K}_e\delta_{ij}{p}^j_{e}V^i_e,  
\end{eqnarray}
\begin{eqnarray}\label{Kdef2}
 \mathcal{K}_e^{\ e_s}&:=&\mathcal{K}_e\delta_{ij}p^j_{e_s}V^i_e, 
\end{eqnarray}
and 
\begin{eqnarray}\label{Kdef3}
  \mathcal{K}_e^{\ e_t}&:=&-\mathcal{K}_e\delta_{ij}p^j_{e_t}\tilde{V}^i_e,  
\end{eqnarray}
where  $e_s$ and $e_t$  satisfy $s(e)=s(e_s)$  and  $t(e)=s(e_t)$, and they are contained in the minimal loop which is used to define $\varrho_e$.

It is necessary to emphasis some points about the densitized extrinsic curvatures on the graph $\gamma$. First, one should  notice that the definition of $\mathcal{K}_e$ relies on the minimal loop since it contains $\varrho_e=\varrho_{e,\square_e}$. Thus, the definitions \eqref{Kdef1},  \eqref{Kdef2} and  \eqref{Kdef3} of  $\mathcal{K}_e^{\ e}$, $\mathcal{K}_e^{\ e_s}$ and $\mathcal{K}_e^{\ e_t}$  on  $\gamma$ are  associated to the specific minimal loops, i.e., $\mathcal{K}_e^{\ e}=\mathcal{K}_{e,\square_e}^{\ e}$, $\mathcal{K}_e^{\ e_s}=\mathcal{K}_{e,\square_{e,e_s}}^{\ e_s}$ and $\mathcal{K}_e^{\ e_t}=\mathcal{K}_{e,\square_{e,e_t}}^{\ e_t}$, where 
$\square_e\ni e$, $\square_{e,e_s}\ni e, e_s$ and $\square_{e,e_t}\ni e, e_t$. Second, it is worth to compare the regularized extrinsic curvature 1-form  $\mathcal{K}^i_{e,s(e)}$ or $\mathcal{K}^i_{e,t(e)}$   to the previous one  \eqref{K1form}. One can see that the expression   \eqref{K1form} is rather complicated in terms of the holonomy-flux variables,  and its quantized version \eqref{Kexist}  is really constructed based on the elementary holonomy-flux operators. The new versions   $\mathcal{K}^i_{e,s(e)}$ and $\mathcal{K}^i_{e,t(e)}$   for the regularized extrinsic curvature 1-form are  clear and simple in terms of the twisted geometry. Hence, if one can find the corresponding twisted geometry operators, one could establish the extrinsic curvature  operators in the quantized twisted geometry based on the expressions   $\mathcal{K}^i_{e,s(e)}$ and $\mathcal{K}^i_{e,t(e)}$  . In the next section of this article, a quantum theory for the twisted geometry will be established in its gauge invariant subspace.

\subsection{Reduced twisted geometry variables }\label{sec:303}
Though we have the twisted geometry to parametrize the holonomy-flux phase space, the gauge degrees of freedom and geometric degrees of freedom are not yet separated clearly. In other words, it is still necessary to find the reduced twisted geometry variables which parametrize the reduced phase space ${H}^+_\gamma$.

\begin{figure}[h]
 \centering
 \includegraphics[scale=0.15]{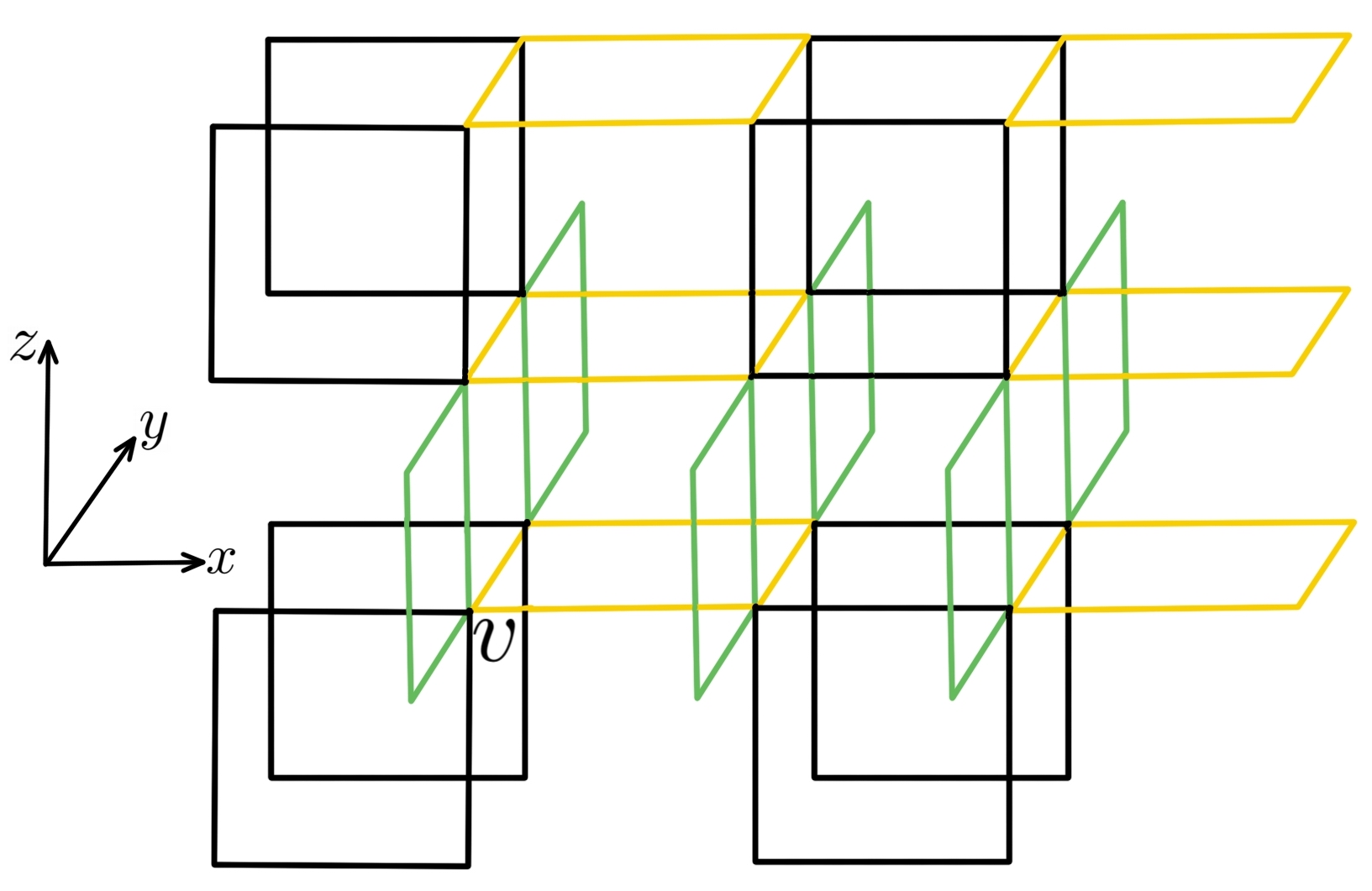}
\caption{The cubic graph on $\mathbb{T}^3$ can be regarded as a composition of square loops, with each edge belonging and only belonging to a square loop. For instance, the 6 edges linked to $v$ are assigned to 3 loops, which are marked with green, yellow and black respectively.}
\label{fig:label1}
\end{figure}

{Recall that the definition of $\zeta_e=\zeta_{e,\square_e}$ depends on a choice of the minimal loop $\square_e$ containing $e$. Indeed, one can fix this choice of the minimal loops for the whole cubic graph $\gamma$ on the manifold with topology $\mathbb{T}^3$. Let us introduce such a choice as follows.} The cubic  graph is adopted to the Cartesian coordinate system $\{x,y,z\}$, with the edges linked to $v$ in this graph being denoted by $e(v,\pm x)$, $e(v,\pm y)$ and $e(v,\pm z)$. Specifically,  $e(v,+x)$ represents the elementary edge started at $v$ along $x$ direction,  $e(v,-x)$ represents the elementary edge ended at $v$ along $x$ direction, and likewise for $e(v,\pm y)$ and $e(v,\pm z)$. Now, one can fix the choice of  the minimal loops $\square_e$ for the definition of  $\zeta_e=\zeta_{e,\square_e}$ on the whole cubic graph, as shown in Fig.\ref{fig:label1} .
For instance,  the six edges linked to the vertex $v$ in Fig.\ref{fig:label1}  belong to the following minimal loops
  \begin{equation}
   \square_{e(v,+y)}=\square_{e(v,+x)}\supset \{e(v,+x), e(v,+y)^{-1}\},
\end{equation}
 \begin{equation}
  \square_{e(v,-z)}=\square_{e(v,-x)}\supset \{e(v,-x), e(v,-z)^{-1}\},
\end{equation}
and 
 \begin{equation}
\square_{e(v,+z)}=  \square_{e(v,-y)}\supset \{e(v,-y), e(v,+z)\}.
\end{equation}
To simplify the notations, let us  denote $e_1(v)=e(v,+x)$, $e_2(v)=e(v,+y)$, $e_3(v)=e(v,-x)$, $e_4(v)=e(v,-z)$, $e_5(v)=e(v,-y)$, $e_6(v)=e(v,+z)$. Then, we can introduce the new notations  $\mathcal{V}^i_{e(v)}\equiv {V}^i_{e(v)}$ for the edge $e$ satisfying $s(e)=v$ and $\mathcal{V}^i_{e(v)}\equiv \tilde{V}^i_{e(v)}$ for the edge $e$ satisfying $t(e)=v$ .  
\begin{figure}[htb]
	\centering
	\begin{tikzpicture} [scale=1.3]

\coordinate  (A) at (-0.1,1.5);
\coordinate  (B) at (0.8,3.5);
\coordinate  (C) at (2.9,4);
\coordinate  (D) at (4.6,3);
\coordinate  (E) at (5,0.8);
\coordinate  (F) at (3.2,-0.9);
\coordinate  (G) at (1.2,-0.5);
\coordinate  (H) at (5.2,-1.3);
\node[scale=0.7] at (A) {$\bullet$} ;
\node[scale=0.7] at (B) {$\bullet$};
\node[scale=0.7] at (C) {$\bullet$};
\node[scale=0.7] at (D) {$\bullet$};
\node[scale=0.7] at (E) {$\bullet$};
\node[scale=0.7] at (F) {$\bullet$};
\node[scale=0.7] at (G) {$\bullet$};
\draw[line width=1pt, ->] (B)  --  node[midway,left=0.1] {$\eta_{e_5(v)}\mathcal{V}^i_{e_5(v)}$} (A) ;
\draw[line width=1pt, ->] (C)  --  node[midway,above=0.2] {$\eta_{e_4(v)}\mathcal{V}^i_{e_4(v)}$} (B) ;
\draw[line width=1pt, ->] (D)  --  node[midway,right=0.1] {$\eta_{e_3(v)}\mathcal{V}^i_{e_3(v)}$} (C) ;
\draw[line width=1pt, ->] (E)  --  node[midway,right=0.1] {$\eta_{e_2(v)}\mathcal{V}^i_{e_2(v)}$} (D) ;
\draw[line width=1pt, ->] (F)  --  node[midway,right=0.15] {$\eta_{e_1(v)}\mathcal{V}^i_{e_1(v)}$} (E) ;
\draw[line width=1pt, ->] (A)  --  node[midway,left=0.1] {$\eta_{e_6(v)}\mathcal{V}^i_{e_6(v)}$} (G) ;
\draw[line width=1pt, ->,yellow] (F)  --  node[midway,below=0.1,black] {$\breve{\eta}_v\mathcal{V}_{v}^i$} (G) ;
\draw[line width=1pt, ->,green] (F)  --  node[midway,left=0.02,black] {$\vec{\eta}_{v,4}$} (B) ;
\draw[line width=1pt, ->,green] (D)  --  node[midway,above=0.02,black] {$\vec{\eta}_{v,2}$} (B) ;
\draw[line width=1pt, ->,green] (F)  --  node[midway,left=0.02,black] {$\vec{\eta}_{v,1}$} (D) ;
\draw[line width=1pt, ->,green] (B)  --  node[midway,left=0.02,black] {$\vec{\eta}_{v,3}$} (G) ;
\draw[line width=1pt, dashed, thick ,red] (F)  --  node[midway,above=0.12,black] {$\  \ \breve{\varrho}_v$}node[midway,above=-0.25,blue] {\scalebox{2}{ $\curvearrowleft$} }(H) ;

\end{tikzpicture}
\caption{The twisted geometry variables associated to the vertex $v$. The blue arrow labeled by the angle $\breve{\varrho}_v$ shows the rotational degrees of freedom around the red dashed line.}
\label{fig:label2}
\end{figure}
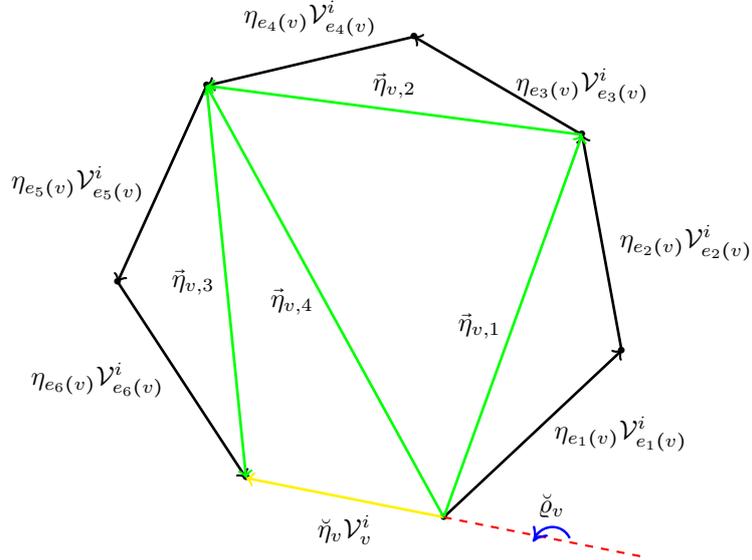


Now, it is ready to consider the reduced twisted geometry variables which parametrizes the $H_\gamma^+$ \cite{Long:2024ldi}. We first have the pair of reduced variables $(\varrho_e,\eta_{e'})$, 
which satisfy
 \begin{equation}\label{repoi1}
 \{\varrho_e,\eta_{e'}\}=\delta_{e,e'} \frac{\kappa}{a^2},\quad {\{\varrho_e,G_v\}=\{\eta_e,G_v\}=0}.
\end{equation}
Then, the unit normal vectors $(\mathcal{V}^i_{e_1(v)},...,\mathcal{V}^i_{e_6(v)})$ can be re-parametrized by  the action-angle variables as follows. 
Let us define 
  \begin{eqnarray}
 \vec{ \eta}_{v,1}\equiv  { \eta}^i_{v,1}&:=&\eta_{e_1(v)}\mathcal{V}^i_{e_1(v)}+\eta_{e_2(v)}\mathcal{V}^i_{e_2(v)},
\end{eqnarray}
 \begin{eqnarray}
 \vec{ \eta}_{v,2}\equiv  { \eta}^i_{v,2}&:=&\eta_{e_3(v)}\mathcal{V}^i_{e_3(v)}+\eta_{e_4(v)}\mathcal{V}^i_{e_4(v)},
\end{eqnarray}
 \begin{eqnarray}
 \vec{ \eta}_{v,3}\equiv  { \eta}^i_{v,3}&:=&\eta_{e_5(v)}\mathcal{V}^i_{e_5(v)}+\eta_{e_6(v)}\mathcal{V}^i_{e_6(v)},
\end{eqnarray}
 \begin{eqnarray}
 \vec{ \eta}_{v,4}\equiv  { \eta}^i_{v,4}&:=&{ \eta}^i_{v,1}+{ \eta}^i_{v,2},
\end{eqnarray}
and 
\begin{equation}
{\eta}_{v,I}=\sqrt{\delta_{ij}{\eta}_{v,I}^i{\eta}_{v,I}^j},\ \text{for}\ I=1,2,3,4.
\end{equation}
In addition, we can define the angles $(\varrho_{v,1},\varrho_{v,2},\varrho_{v,3},\varrho_{v,4})$ between the triangles  in Fig.\ref{fig:label2}. Specifically, we define  \cite{Long:2024ldi,PhysRevD.83.044035}:
\begin{itemize}
    \item[(1)] $\varrho_{v,1}$  as the angle between the plane identified by the vectors $ (\mathcal{V}^i_{e_1(v)},\mathcal{V}^i_{e_2(v)})$ and the plane identified by the vectors $( \vec{ \eta}_{v,1},\vec{ \eta}_{v,2})$;
    \item[(2)]  $\varrho_{v,2}$  as the angle between the plane identified by the vectors $ (\mathcal{V}^i_{e_3(v)},\mathcal{V}^i_{e_4(v)})$ and the plane identified by the vectors $( \vec{ \eta}_{v,1},  \vec{ \eta}_{v,2})$;
    \item[(3)]  $\varrho_{v,3}$  as the angle between the plane identified by the vectors $ (\mathcal{V}^i_{e_5(v)},\mathcal{V}^i_{e_6(v)})$ and the plane identified by the vectors $( \vec{ \eta}_{v,3},  \vec{ \eta}_{v,4})$; and
    \item[(4)] $\varrho_{v,4}$  as the angle between the plane identified by the vectors $( \vec{ \eta}_{v,1},  \vec{ \eta}_{v,2})$   and the plane identified by the vectors $( \vec{ \eta}_{v,3},  \vec{ \eta}_{v,4})$. 
\end{itemize}
Moreover, we define
\begin{equation}
\mathcal{V}_{v}\equiv\mathcal{V}_{v}^i\tau_i:=\frac{G^i_v\tau_i}{\breve{\eta}_v}
\end{equation}
with  $\breve{\eta}_v:=\sqrt{\delta_{ij}G^i_vG^j_v}$ being the module of the Gauss constraint.
{Then, an extra angle variable $\breve{\varrho}_v$ can be defined as the angle between the plane identified by  the $su(2)$-valued vectors $({ \eta}^i_{v,3}\tau_i, \mathcal{V}_{v}^i\tau_i)$ and the plane identified by  $(n(\mathcal{V}_{v})\tau_2 n(\mathcal{V}_{v})^{-1}, \mathcal{V}_{v}^i\tau_i ) $,  see the illustration of this definition in Fig.\ref{fig:label2}.
It is directly to check that  $\breve{\varrho}_v$  is conjugate to the module $\breve{\eta}_v$ of Gauss constraint, i.e.,
  \begin{equation}
 \{\breve{\eta}_v,\breve{\varrho}_{v'}\}=\delta_{v,v'}\frac{\kappa}{a^2}.
\end{equation}
}

It is necessary to introduce the interior $\dot{P}_\gamma^+$ of the twisted geometry space $P_\gamma^+$, with $\dot{P}_\gamma^+$ being composed by  the phase space points satisfying the following conditions,
 \begin{eqnarray}\label{boundary000}
 |\eta_{e_1(v)}-\eta_{e_2(v)}|<{ \eta}_{v,1}<\eta_{e_1(v)}+\eta_{e_2(v)},&&\ 
  \  |\eta_{e_3(v)}-\eta_{e_4(v)}|<{ \eta}_{v,2}<\eta_{e_3(v)}+\eta_{e_4(v)},\\\nonumber|\eta_{e_5(v)}-\eta_{e_6(v)}|<{ \eta}_{v,3}<\eta_{e_5(v)}+\eta_{e_6(v)},&&\  \  |{ \eta}_{v,1}-{ \eta}_{v,2}|<{ \eta}_{v,4}<{ \eta}_{v,1}+{ \eta}_{v,2},\\\nonumber
 |{ \eta}_{v,3}-\breve{ \eta}_{v}|<{ \eta}_{v,4}<{ \eta}_{v,3}+\breve{ \eta}_{v},&&\quad \breve{ \eta}_{v}> 0,{ \eta}_{e}> 0,
 \forall v\in V(\gamma), e\in E(\gamma).
\end{eqnarray}
One can notice that the  inequalities in  \eqref{boundary000} determine the non-vanishing triangles in Fig.\ref{fig:label2}. Moreover, some of the triangles in  Fig.\ref{fig:label2} are vanished on the boundary of $\dot{P}_\gamma^+$, and the vanishing triangles also reduce the degrees of freedom of the corresponding angle variables \cite{PhysRevLett.107.011301,Bianchi:2012wb}.
Now, it is easy to see that the unit vectors $(\mathcal{V}^i_{e_1(v)},...,\mathcal{V}^i_{e_6(v)})$  in $\dot{P}_\gamma^+$ can be determined by $(\breve{\eta}_v,\breve{\varrho}_{v'})$, $\mathcal{V}_v$ and  $({ \eta}_{v,I}, \varrho_{v',I}), I\in\{1,2,3,4\}$ uniquely. As a summary, one can conclude that the new twisted geometric variables   $$(\eta_e,\varrho_e),({ \eta}_{v,I}, \varrho_{v,I})|_{I=1,2,3,4}, (\breve{\eta}_v,\breve{\varrho}_{v}), \mathcal{V}_v$$provide a re-parametrization of the interior $\dot{P}_\gamma^+$  of the twisted geometry space $P_\gamma^+$.

\begin{figure}[htb]
	\centering
	\begin{tikzpicture} [scale=1.3]

\coordinate  (A) at (0.2,1);

\coordinate  (B) at (0.35,3.2);
\coordinate  (C) at (2.5,4.2);

\coordinate  (D) at (4.5,3);
\coordinate  (E) at (4.5,0.5);

\coordinate  (F) at (2.4,-0.5);

\node[scale=0.7] at (A) {$\bullet$} ;
\node[scale=0.7] at (B) {$\bullet$};
\node[scale=0.7] at (C) {$\bullet$};
\node[scale=0.7] at (D) {$\bullet$};
\node[scale=0.7] at (E) {$\bullet$};
\node[scale=0.7] at (F) {$\bullet$};
\draw[line width=1pt, ->] (B)  --  node[midway,left=0.1] {$\eta_{e_5(v)}\mathcal{V}^i_{e_5(v)}$} (A) ;
\draw[line width=1pt, ->] (C)  --  node[midway,left=0.2] {$\eta_{e_4(v)}\mathcal{V}^i_{e_4(v)}$} (B) ;
\draw[line width=1pt, ->] (D)  --  node[midway,right=0.1] {$\eta_{e_3(v)}\mathcal{V}^i_{e_3(v)}$} (C) ;
\draw[line width=1pt, ->] (E)  --  node[midway,right=0.1] {$\eta_{e_2(v)}\mathcal{V}^i_{e_2(v)}$} (D) ;
\draw[line width=1pt, ->] (F)  --  node[midway,right=0.15] {$\eta_{e_1(v)}\mathcal{V}^i_{e_1(v)}$} (E) ;
\draw[line width=1pt, ->] (A)  --  node[midway,left=0.1] {$\eta_{e_6(v)}\mathcal{V}^i_{e_6(v)}$} (F) ;
\draw[line width=1pt, ->,green] (B)  --  node[midway,left=0.02,black] {$\vec{\eta}_{v,3}$} (F) ;
\draw[line width=1pt, ->,green] (D)  --  node[midway,above=0.02,black] {$\vec{\eta}_{v,2}$} (B) ;
\draw[line width=1pt, ->,green] (F)  --  node[midway,left=0.02,black] {$\vec{\eta}_{v,1}$} (D) ;

\end{tikzpicture}
\caption{The reduced twisted geometry associated to the vertex $v$.}
\label{fig:label222}
\end{figure}
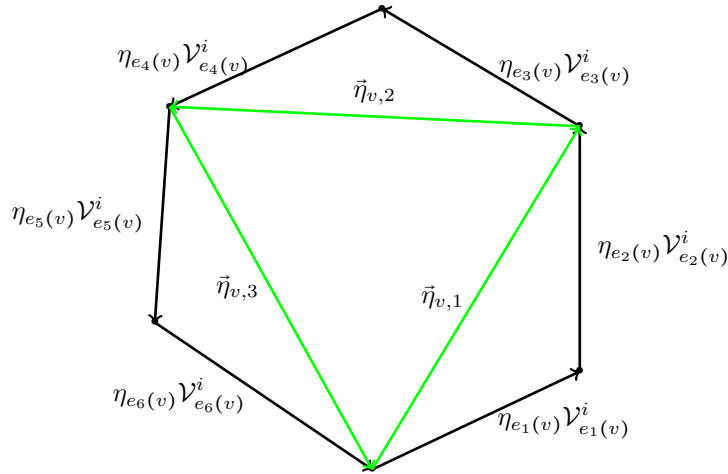

The gauge reduction with respect to the Gauss constraint in $P_\gamma^+$ can be carried out by imposing the condition $\breve{\eta}_v=0$.  In fact, the reduced twisted geometry space $H_\gamma^+$ is given by taking the limit $\breve{\eta}_v\to0$ in $P_\gamma^+$ for all $v\in \gamma$, which vanishes the triangles spanned by $\breve{ \eta}_{v}, { \eta}_{v,3}$ and $ { \eta}_{v,4}$ in Fig.\ref{fig:label2}. Then, one can obtain the reduced twisted geometry space  $H_\gamma^+$, which is parametrized by the reduced twisted geometric variables   $(\eta_e,\varrho_e),({ \eta}_{v,I}, \varrho_{v,I})|_{I=1,2,3} $; Here one should notice that  $\varrho_{v,3}$ in $H_\gamma^+$  is  the angle between the plane identified by the vectors $ (\mathcal{V}^i_{e_5(v)},\mathcal{V}^i_{e_6(v)})$ and the plane identified by the vectors $( \vec{ \eta}_{v,1},  \vec{ \eta}_{v,2})$, as shown in Fig.\ref{fig:label222}. One can also introduce the interior $\dot{H}_\gamma^+$ of the reduced twisted geometry space $H_\gamma^+$, with $\dot{H}_\gamma^+$ being composed by  the phase space points satisfying the following conditions,
 \begin{eqnarray}\label{boundary}
&& |\eta_{e_1(v)}-\eta_{e_2(v)}|<{ \eta}_{v,1}<\eta_{e_1(v)}+\eta_{e_2(v)},\ \ \
    |\eta_{e_3(v)}-\eta_{e_4(v)}|<{ \eta}_{v,2}<\eta_{e_3(v)}+\eta_{e_4(v)},\\\nonumber
    &&|\eta_{e_5(v)}-\eta_{e_6(v)}|<{ \eta}_{v,3}<\eta_{e_5(v)}+\eta_{e_6(v)},  \  \ \   |{ \eta}_{v,1}-{ \eta}_{v,2}|<{ \eta}_{v,3}<{ \eta}_{v,1}+{ \eta}_{v,2},\\\nonumber
&& \ \    \eta_e>0,\ 
 \forall v\in V(\gamma), e\in E(\gamma).
\end{eqnarray}
Now, the following theorem shows that the symplectic structure on $\dot{P}_\gamma^+$ and $\dot{H}_\gamma^+$ can be expressed in terms of the new geometric variables \cite{Long:2024ldi}.
\begin{theorem}\label{theorem1}
     For the twisted geometry space $P_\gamma^+=(\times_{e\in E(\gamma)} T^\ast S_e^1)^+\times_{v\in V(\gamma)} \mathfrak{P}_v$ on a the cubic graph $\gamma$ on $\mathbb{T}^3$, where $(\eta_e,\varrho_e)\in (T^\ast S_e^1)^+\cong \mathbb{R}^+_e\times S^1_e$ and $(({ \eta}_{v,I}, \varrho_{v,I}), (\breve{\eta}_v,\breve{\varrho}_{v}), \mathcal{V}_v^i)\in \mathfrak{P}_v, \quad I\in\{1,...,4\}$,
the symplectic 1-form on its interior $\dot{P}_\gamma^+$ of $P_\gamma^+$ can be re-expressed as
  \begin{equation}\label{sym3}
  \Theta_{\dot{P}_\gamma^+}=\frac{a^2}{\kappa}\left(\sum_{e\in\gamma}\eta_ed\varrho_e+\sum_{v\in\gamma}\left(\breve{\eta}_v\text{Tr}(\mathcal{V}_vdn_vn_v^{-1}) +\sum_{I=1}^{4}{ \eta}_{v,I}d\varrho_{v,I} +\breve{\eta}_vd\breve{\varrho}_{v}\right)\right).
\end{equation}
By carrying out the symplectic reduction with respect to the Gauss constraint, the reduced symplectic 1-form in the interior $\dot{H}_\gamma^+$ of the reduced phase space $H_\gamma^+$ can be given by \footnote{  Notice that  $\varrho_{v,3}$ in $H_\gamma^+$  is  the angle between the plane identified by the vectors $ (\mathcal{V}^i_{e_5(v)},\mathcal{V}^i_{e_6(v)})$ and the plane identified by the vectors $( \vec{ \eta}_{v,1},  \vec{ \eta}_{v,2})$, as shown in Fig.\ref{fig:label222}. Thus, one has $  \varrho_{v,3}|_{\dot{H}_\gamma^+}:=\lim_{\breve{\eta}_v\to 0}  (\varrho_{v,3}+\varrho_{v,4})|_{\dot{P}_\gamma^+}$ and $  2\eta_{v,3}|_{\dot{H}_\gamma^+}=\lim_{\breve{\eta}_v\to 0}  (\eta_{v,3}+\eta_{v,4})|_{\dot{P}_\gamma^+}$.  Then, by using $\frac{1}{2}(\eta_{v,3}+\eta_{v,4})d(\varrho_{v,3}+\varrho_{v,4})+\frac{1}{2}(\eta_{v,3}-\eta_{v,4})d(\varrho_{v,3}-\varrho_{v,4})=\eta_{v,3}d\varrho_{v,3}+\eta_{v,4}d\varrho_{v,4}$, the relation between $\Theta_{\dot{H}_\gamma^+}$ and $\Theta_{\dot{P}_\gamma^+}$ can be established as $  \Theta_{\dot{H}_\gamma^+}=\lim_{\breve{\eta}_v\to 0}  \Theta_{\dot{P}_\gamma^+}.$} 
  \begin{equation}\label{sym04}
  \Theta_{\dot{H}_\gamma^+}=\frac{a^2}{\kappa}\left(\sum_{e\in\gamma}\eta_ed\varrho_e+\sum_{v\in\gamma}\sum_{I=1}^{3}{ \eta}_{v,I} d\varrho_{v,I}\right).
\end{equation}
\end{theorem}

It is still necessary to analyze the boundary $\bar{H}_\gamma^+$ of the reduced phase space $\dot{H}_\gamma^+$. As mentioned above, the boundary of $\dot{H}_\gamma^+$ is given by vanishing some of the triangles in Fig.\ref{fig:label222}. 
In fact, there are two methods to vanish a triangle in Fig.\ref{fig:label222}, which lead to two types of vanishing triangles in $\bar{H}_\gamma^+$. The first type vanishing triangle has some vanishing edges, while the second type vanishing triangle has no vanishing edge.
Correspondingly, the  boundary $\bar{H}_\gamma^+$ of the reduced phase space $\dot{H}_\gamma^+$ can be decomposed as 
  \begin{equation}
\bar{H}_\gamma^+=\bar{H}^{'+}_\gamma\cup \bar{H}^{''+}_\gamma,
\end{equation}
where $\bar{H}^{'+}_\gamma$ only contains the first type vanishing triangles, and   $\bar{H}^{''+}_\gamma$ must contain the second type vanishing triangles.
The symplectic structure on $\bar{H}^{'+}_\gamma$ can be obtained by taking the associated limit of $\Theta_{\dot{H}_\gamma^+}$. For example, the symplectic structure on the subspace of $\bar{H}^{'+}_\gamma$ defined by $\lim_{\eta_{e_1(v')}\to0}{\dot{H}_\gamma^+}$ can be given by $\lim_{\eta_{e_1(v')}\to0}\Theta_{\dot{H}_\gamma^+}$, which reads
  \begin{equation}
\lim_{\eta_{e_1(v')}\to0}\Theta_{\dot{H}_\gamma^+}=\frac{a^2}{\kappa}\left(\sum_{e\in E(\gamma)\setminus e_1(v)}\eta_ed\varrho_e+\sum_{v\in V(\gamma)\setminus v'}\sum_{I=1}^{3}{ \eta}_{v,I} d\varrho_{v,I}+\sum_{I=1}^{2}{ \eta}_{v',I} d\varrho_{v',I}\right),
\end{equation}
where we used $  \varrho_{e_2(v')}|_{\eta_{e_1(v')}=0}:=\lim_{\eta_{e_1(v')}\to0}  (\varrho_{v',1}+\varrho_{e_2(v')})|_{\Theta_{\dot{H}_\gamma^+}}$ , $  2 \eta_{e_2(v')}|_{\eta_{e_1(v')}=0}=\lim_{\eta_{e_1(v')}\to0}(\eta_{v',1}+\eta_{e_2(v')})|_{\Theta_{\dot{H}_\gamma^+}}$ and  $\frac{1}{2}(\eta_{v',1}+\eta_{e_2(v')})d(\varrho_{v',1}+\varrho_{e_2(v')})+\frac{1}{2}(\eta_{v',1}-\eta_{e_2(v')})d(\varrho_{v',1}-\varrho_{e_2(v')})=\eta_{v',1}d\varrho_{v',1}+\eta_{e_2(v')}d\varrho_{e_2(v')}$. 
Thus, one can conclude that the space $\dot{H}_\gamma^+\cup \bar{H}^{'+}_\gamma$ is a presymplectic manifold equipped with the  (pre-)symplectic potential
  \begin{equation}\label{sym4}
  \Theta_{\dot{H}_\gamma^+\cup\bar{H}^{'+}_\gamma}=\frac{a^2}{\kappa}\left(\sum_{e\in\gamma}\eta_ed\varrho_e+\sum_{v\in\gamma}\sum_{I=1}^{3}{ \eta}_{v,I} d\varrho_{v,I}\right),
\end{equation}
where $  \Theta_{\dot{H}_\gamma^+\cup\bar{H}^{'+}_\gamma}$ gives a degenerate symplectic 2-form on $\bar{H}^{'+}_\gamma$. Also, one can impose the reduction with respect to the kernel of $\Theta_{\dot{H}_\gamma^+\cup\bar{H}^{'+}_\gamma}$ on $\bar{H}^{'+}_\gamma$, which leads to a symplectic manifold with non-degenerate symplectic 2-form given by $ \Theta_{\dot{H}_\gamma^+\cup\bar{H}^{'+}_\gamma}$ projected on the reduced space.
Additionally, we still need to consider the symplectic structure on $\bar{H}^{''+}_\gamma$. Nevertheless, we will show that the quantum theory is irrelevant to the subspace $\bar{H}^{''+}_\gamma$ of the boundary of $\dot{H}_\gamma^+$ in the next section; thus, we will not analyze the associated symplectic structure on $\bar{H}^{''+}_\gamma$ in this article.


\section{Quantization of reduced twisted geometry}\label{sec4}

It is known that the quantum representation of the holonomy-flux algebra on $\gamma$ gives the Hilbert space $\mathcal{H}_\gamma$ spanned by the spin-network states. Especially, one can impose the quantum Gauss constraint in $\mathcal{H}_\gamma$ , which leads to the gauge invariant Hilbert space   $\mathcal{H}^{\text{inv}}_\gamma$. Notice that the reduced twisted geometry is just the reduced classical holonomy-flux with respect to Gauss constraint. Hence, it is expected that the quantum representation of some Poisson algebras in the reduced twisted geometry on $\gamma$ gives the  gauge invariant Hilbert space $\mathcal{H}^{\text{inv}}_\gamma$. In the following part of this section, we will introduce a regularization scheme for the reduced twisted geometry variables, and show that the representation of the Poisson algebra among these regularized variables gives the  gauge invariant Hilbert space   $\mathcal{H}^{\text{inv}}_\gamma$ exactly.

\subsection{Quantum algebra of reduced geometry variables}\label{sec:401}

As mentioned above,  our goal is to find the specific Poisson brackets in the reduced twisted geometry space on $\gamma$, whose quantum representation gives the  gauge invariant Hilbert space   $\mathcal{H}^{\text{inv}}_\gamma$.  In other words,  we need to find the fundamental algebras of the Hilbert space  $\mathcal{H}^{\text{inv}}_\gamma$ based on the reduced twisted geometry.

Before turning to the fundamental algebra of the Hilbert space $\mathcal{H}^{\text{inv}}_\gamma$, let us first introduce some details of the spaces $\mathcal{H}_\gamma$ and $\mathcal{H}^{\text{inv}}_\gamma$.
 The Hilbert space $\mathcal{H}_{\gamma}$ on $\gamma$  is composed of the square-integrable functions on $SU(2)$ associated to each edge $e\in E(\gamma)$.
 Specifically, a square integrable functions on $\gamma$ takes the formulation
\begin{equation}
\Psi_\gamma=\Psi_\gamma(\{h_{e}\}_{e\in E(\gamma)}).
\end{equation}
The $SU(2)$ gauge transformation of $\Psi_\gamma$ reads
\begin{equation}
\Psi_\gamma(\{h_{e}\}_{e\in E(\gamma)})\to\Psi_\gamma(\{g_{s(e)}h_{e}g^{-1}_{t(e)}\}_{e\in E(\gamma)}),
\end{equation}
where $\{g_v|v\in V(\gamma)\}$ are given at each vertex $v\in V(\gamma)$ respectively, $s(e)$ represents the source vertex of $e$ and
$t(e)$ the target vertex of $e$. The spin-network states provide a basis of space $\mathcal{H}_\gamma$. Specifically, a spin-network basis state on $\gamma$ is given by labeling a spin $j_e\in\frac{\mathbb{N}}{2}$ on each edge $e\in E(\gamma)$ and an intertwiner $\check{\mathcal{I}}_v$ on each vertex $v\in V(\gamma)$, which reads \cite{Ashtekar:2004eh} 
\begin{equation}\label{spinbasis}
\Psi_{\gamma,\{j_e,\mathcal{I}_v\}}=\text{tr}\left(\bigotimes_{e\in E(\gamma)}\pi_{j_e}(h_e)\bigotimes_{v\in V(\gamma)} \check{\mathcal{I}}_v \right)
\end{equation}
where $\pi_{j_e}(h_e)$ is the representation matrix of $h_e\in SU(2)$ in the representation space $V^{j_e}$ of $SU(2)$ labelled by spin $j_e$. 

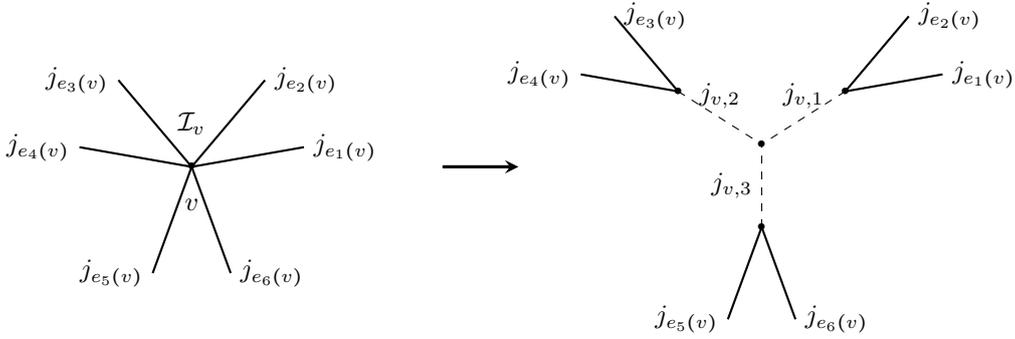
\begin{figure}[htb]
	\centering
	\begin{tikzpicture} [scale=1]
\coordinate  (O) at (0,0);

\coordinate  (O1) at (-1,0.3);

\coordinate  (O2) at (-1,-0.3);

\coordinate  (O3) at (1,0.3);

\coordinate  (O4) at (1,-0.3);

\coordinate  (A) at (6.4,1);

\coordinate  (A1) at (5,0.3);

\coordinate  (A2) at (5,-0.3);

\coordinate  (B) at (8.6,1);

\coordinate  (C) at (7.5,-0.8);

\coordinate  (D) at (7.5,0.3);

\coordinate  (B1) at (10,0.3);

\coordinate  (B2) at (10,-0.3);

\node[scale=0.7] at (O) {$\bullet$} node[below=0.3] {$v$}node [above=0.3] {$\mathcal{I}_v$};

\draw[thick] (O)  --  node[midway,sloped] {}++ (50:1.5) node[right] {$j_{e_2(v)}$};

\draw[thick] (O)  to  node[midway,sloped]{} ++ (10:1.5) node[right] {$j_{e_1(v)}$};

\draw[thick] (O)  --  node[midway,sloped]{} ++ (-70:1.5) node[right] {$j_{e_6(v)}$};

\draw[thick] (O)  to  node[midway,sloped]{} ++ (-110:1.5) node[left] {$j_{e_5(v)}$};

\draw[thick] (O)  to  node[midway,sloped]{} ++ (130:1.5) node[left] {$j_{e_3(v)}$};

\draw[thick] (O)  to  node[midway,sloped]{} ++ (170:1.5) node[left]{$j_{e_4(v)}$} ;




\draw[->,>=stealth,very thick] (O) ++ (3.3,0) -- ++ (1,0);



\draw[thick] (B)  --  node[midway,sloped]{} ++ (50:1.3) node[right] {$j_{e_2(v)}$} ;

\draw[thick] (B)  to  node[midway,sloped]{} ++ (10:1.3) node[right]{$j_{e_1(v)}$};
\draw[thick] (A)  to  node[midway,sloped]{} ++ (130:1.3) node[right]{$j_{e_3(v)}$};

\draw[thick] (A)  to  node[midway,sloped]{} ++ (170:1.3) node[left]{$j_{e_4(v)}$};



\draw[thick] (C)  to  node[midway,sloped]{} ++ (-70:1.3) node[right]{$j_{e_6(v)}$};
\draw[thick] (C)  to  node[midway,sloped]{} ++ (-110:1.3) node[left]{$j_{e_5(v)}$};

\draw[dashed] (A) -- node[midway,sloped]{} node[above,black] {$j_{v,2}$} (D);
\draw[dashed] (B) -- node[midway,sloped]{} node[above,black] {$j_{v,1}$} (D);
\draw[dashed] (C) -- node[midway,sloped]{} node[left,black] {$j_{v,3}$} (D);

\node[scale=0.7] at (A) {$\bullet$};
\node[scale=0.7] at (B) {$\bullet$};
\node[scale=0.7] at (C) {$\bullet$};
\node[scale=0.7] at (D) {$\bullet$};
\end{tikzpicture}
\caption{The illustration of recoupling scheme for $\mathcal{I}_v=\mathcal{I}_{v,\{j_{v,1},j_{v,2},j_{v,3}\}}$.}
\label{figrecouple}
\end{figure}

The spin-network state $\Psi_{\gamma,\{j_e,\mathcal{I}_v\}}$ is $SU(2)$ gauge invariant if and only if each $v\in V(\gamma)$ is labeled by a gauge invariant intertwiner $\mathcal{I}_v$. More explicitly, the gauge invariant intertwiner $\mathcal{I}_v$ at vertex $v$ is a $SU(2)$-invariant state  in the tensor product space of all the spins associated to the edges linked to $v$,
\begin{equation}
\mathcal{I}_v\in\mathcal{H}_v^{\{j_e\}}:= \text{Inv}_{SU(2)}\left[\bigotimes_{e|t(e)=v} V^{j_e}
\otimes
\bigotimes_{e|s(e)=v} \bar{V}^{j_e}\right],
\end{equation}
where $\bar{V}^{j}$ is the dual space of $V^{j}$. 
An orthonormal basis of the intertwiner space $\mathcal{H}_v^{\{j_e\}}$ can be established by the recoupling scheme. Let us consider the recoupling scheme which is adapted to the reduced twisted geometry variables at each vertex $v$. Specifically, for each 6-valents vertex in the cubic graph, one can  introduce three internal edges as shown in Fig.\ref{figrecouple}. Then, with the spins $\{j_e\in\frac{\mathbb{N}}{2}|e\in E(\gamma)\}$ labeled to each edges being given, an orthonormal basis of the gauge invariant  intertwiner space $\mathcal{H}_v^{\{j_e\}}$ is given by
\begin{equation}
\{\mathcal{I}_{v,\{j_{v,1},j_{v,2},j_{v,3}\}}\in \mathcal{H}_v^{\{j_e\}}\}.
\end{equation}
Especially, the internal spins  $\{j_{v,1},j_{v,2},j_{v,3}\}$  labels the three internal edges associated to $v$ in the recoupling scheme as shown in Fig.\ref{figrecouple}, which are restricted by
 \begin{eqnarray}\label{quanboundary}
 |j_{e_1(v)}-j_{e_2(v)}|\leq { j}_{v,1}\leq j_{e_1(v)}+j_{e_2(v)},&&\ 
  \  |j_{e_3(v)}-j_{e_4(v)}|\leq{ j}_{v,2}\leq j_{e_3(v)}+j_{e_4(v)},\\\nonumber|j_{e_5(v)}-j_{e_6(v)}|\leq { j}_{v,3}\leq j_{e_5(v)}+j_{e_6(v)},&&\  \  |{ j}_{v,1}-{ j}_{v,2}|\leq { j}_{v,3}\leq{ j}_{v,1}+{ j}_{v,2}.
\end{eqnarray}
It is worth to emphasize that Eq.\eqref{quanboundary} only restricts the relations among the internal spins $\{j_{v,1},j_{v,2},j_{v,3}\}$ and the edge spins $j_e$, with all of them taking values in the range $\frac{\mathbb{N}}{2}$.  Let us denote the gauge invariant spin-network state $\Psi_{\gamma,\{j_e,\mathcal{I}_v\}}$ with $\mathcal{I}_{v}=\mathcal{I}_{v,\{j_{v,1},j_{v,2},j_{v,3}\}}$ by $|\vec{j},\vec{\mathfrak{I}}\rangle$, where $\vec{j}=\{...,j_e,...\}|_{e\in E(\gamma)}$, $\vec{\mathfrak{I}}=\{...,\mathfrak{I}_v,...\}  |_{v\in V(\gamma)}$ with $\mathfrak{I}_v=\{j_{v,1},j_{v,2},j_{v,3}\}$. Moreover, one can conclude that the gauge invariant state $|\vec{j},\vec{\mathfrak{I}}\rangle$ provides a complete and orthogonal basis of the gauge invariant subspace  $\mathcal{H}_\gamma^{\text{inv}}$ of  $\mathcal{H}_\gamma$.

Before we use the symplectic 1-form \eqref{sym4} to study the fundamental quantum algebra for the Hilbert space $\mathcal{H}_\gamma^{\text{inv}}$, it is necessary to emphasize that the symplectic 1-form \eqref{sym4} on the twisted geometry space is only valid in the space $\dot{H}_\gamma^+\cup\bar{H}^{'+}_\gamma$. Hence, we need to consider the quantum version of the condition \eqref{boundary} and the first type vanishing triangles in Fig.\ref{fig:label222}.
Recalling the relation \eqref{para} between the fluxes and the twisted geometry variables, one can immediately give the well-defined operators $\hat{\eta}_{v,1}$, $\hat{\eta}_{v,2}$, $\hat{\eta}_{v,3}$ and $\hat{\eta}_{e}$ acting in $\mathcal{H}_\gamma^{\text{inv}}$. Specifically, the basis state $|\vec{j},\vec{\mathfrak{I}}\rangle$ are the common eigen-states of the operators   $\hat{\eta}_{v,1}$, $\hat{\eta}_{v,2}$, $\hat{\eta}_{v,3}$ and  $\hat{\eta}_{e}$, with the eigen-equations reading
\begin{equation}
\hat{\eta}_{e}|\vec{j},\vec{\mathfrak{I}}\rangle=t\sqrt{j_e(j_e+1)}|\vec{j},\vec{\mathfrak{I}}\rangle,\quad \hat{\eta}_{v,I}|\vec{j},\vec{\mathfrak{I}}\rangle=t\sqrt{j_{v,I}(j_{v,I}+1)}|\vec{j},\vec{\mathfrak{I}}\rangle,\  \text{for}\ I=1,2,3.
\end{equation}
Based on these eigen-equations, it is immediately to give the quantum version of the condition \eqref{boundary} in the  basis $|\vec{j},\vec{\mathfrak{I}}\rangle$, which reads 
 \begin{eqnarray}\label{boundaryquantum00}
 |\sqrt{j_{e_1(v)}(j_{e_1(v)}+1)}-\sqrt{j_{e_2(v)}(j_{e_2(v)}+1)}|&<&\sqrt{j_{v,1}(j_{v,1}+1)}\\\nonumber
&<&\sqrt{j_{e_1(v)}(j_{e_1(v)}+1)}+\sqrt{j_{e_2(v)}(j_{e_2(v)}+1)},\\\nonumber
 \ 
  \  |\sqrt{j_{e_3(v)}(j_{e_3(v)}+1)}-\sqrt{j_{e_4(v)}(j_{e_4(v)}+1)}|&<&\sqrt{j_{v,2}(j_{v,2}+1)}\\\nonumber
&<&\sqrt{j_{e_3(v)}(j_{e_3(v)}+1)}+\sqrt{j_{e_4(v)}(j_{e_4(v)}+1)},\\\nonumber
 |\sqrt{j_{e_5(v)}(j_{e_5(v)}+1)}-\sqrt{j_{e_6(v)}(j_{e_6(v)}+1)}|&<&\sqrt{j_{v,3}(j_{v,3}+1)}\\\nonumber
&<&\sqrt{j_{e_5(v)}(j_{e_5(v)}+1)}+\sqrt{j_{e_6(v)}(j_{e_6(v)}+1)},\\\nonumber
 |\sqrt{j_{v,1}(j_{v,1}+1)}-\sqrt{j_{v,2}(j_{v,2}+1)}|&<&\sqrt{j_{v,3}(j_{v,3}+1)}\\\nonumber
&<&\sqrt{j_{v,1}(j_{v,1}+1)}+\sqrt{j_{v,2}(j_{v,2}+1)},
\\\nonumber
 j_e>0,\quad \forall e\in E(\gamma), \quad  \forall v\in V(\gamma).&&
\end{eqnarray}
Recalling that the the  basis $|\vec{j},\vec{\mathfrak{I}}\rangle$ of $\mathcal{H}_\gamma^{\text{inv}}$  satisfies \eqref{quanboundary}, one can  see that the quantum version \eqref{boundaryquantum00} of the condition \eqref{boundary}  hold  in the subspace $\tilde{\mathcal{H}}_\gamma^{\text{inv}}$ defined by
\begin{equation}
\tilde{\mathcal{H}}_\gamma^{\text{inv}}:={\mathcal{H}}_\gamma^{\text{inv}}|_{j_e,j_{v,I}\neq 0,\forall e\in E(\gamma), v\in V(\gamma),I\in\{1,2,3\}}.
\end{equation}
In fact, the quantum version \eqref{boundaryquantum00} of the condition \eqref{boundary}  ensures that the quantized triangles in Fig.\ref{fig:label222} are non-vanished in space $\tilde{\mathcal{H}}_\gamma^{\text{inv}}$. Moreover,  some of the quantized triangles in Fig.\ref{fig:label222} are vanished for the states in the space ${\mathcal{H}}_\gamma^{\text{inv}}\setminus \tilde{\mathcal{H}}_\gamma^{\text{inv}}$, and these vanishing triangles must be the first-type vanishing triangles. Now, recalling that the subspace $\bar{H}^{''+}_\gamma$ of the boundary of the reduced phase space $H_\gamma^+$ must contain the second type
vanishing triangles, one can conclude that the gauge invarinat Hilbert space ${\mathcal{H}}_\gamma^{\text{inv}}$  is irrelevant to $\bar{H}^{''+}_\gamma$. Hence,  ${\mathcal{H}}_\gamma^{\text{inv}}$ can be regarded as the quantum representation space of some fundamental algebras in the subspace $\dot{H}_\gamma^+\cup \bar{H}^{'+}_\gamma$ of the reduced phase space $H_\gamma^+$. 

Now, let us start to consider the fundamental algebra in the space  $\dot{H}_\gamma^+\cup \bar{H}^{'+}_\gamma$  which generates the Hilbert space ${\mathcal{H}}_\gamma^{\text{inv}}$ based on  the symplectic 1-form \eqref{sym4}.
Recall the Poisson algebra given by  the symplectic 1-form \eqref{sym4}, for which the Poisson brackets read
\begin{eqnarray}\label{Pochi2}
&&\{\varrho_e,{\varrho}_{e'}\}=\{\varrho_e,{\varrho}_{v,I}\}=\{\varrho_{v',I'},{\varrho}_{v,I}\}=\{\eta_{e'},{ \eta}_{e}\}=\{\eta_{v,I},{ \eta}_{v',I'}\}=\{\eta_{e},{ \eta}_{v,I}\}=0,\\\nonumber
&&\{\varrho_e,\eta_{e'}\}=\delta_{e,e'}\frac{\kappa}{a^2},\quad \{{\varrho}_{v,I},{ \eta}_{v',I'}\}=\delta_{v,v'}\delta_{I,I'}\frac{\kappa}{a^2}.
\end{eqnarray}
{It is easy to see that there is no operator corresponding to $\varrho_e$ and ${\varrho}_{v,I}$ on ${\mathcal{H}}_\gamma^{\text{inv}}$  compatiable with the above Poisson brackets, as the equations $[\varrho,\eta] |{j},{\mathfrak{I}}\rangle =i \hbar \frac{\kappa}{a^2}|{j},{\mathfrak{I}}\rangle$ does not have a solution.}
Thus, the fundamental variables for  ${\mathcal{H}}_\gamma^{\text{inv}}$  should  be some  regularized version of $\varrho_e,\eta_{e'},{\varrho}_{v,I},{ \eta}_{v',I'}$. {Inspired by the quantization procedure of loop quantum gravity, we expect the quantization to be taken here will be similar to the polymer quantization.} In other words, we want to find a canonical transformation of the Poisson brackets \eqref{Pochi2}, so that one can establish the quantum representation of the resulting Poisson brackets of the canonical transformation in the Hilbert space ${\mathcal{H}}_\gamma^{\text{inv}}$ . Such a canonical transformation can be constructed by the following three steps.
\begin{enumerate}
    \item We first notice that the complete and orthogonal basis   $|\vec{j},\vec{\mathfrak{I}}\rangle$  of the Hilbert space ${\mathcal{H}}_\gamma^{\text{inv}}$  are labelled by the quantum numbers $j_e$ and $j_{v,I}$. Also, one can define two sets of the operators   $\hat{\chi}_{e}$ and $\hat{\chi}_{v,I}$  as
\begin{equation}
\hat{\chi}_e:=\sqrt{(\hat{\eta}_{e}/t)^2+1/4}-1/2,\quad \hat{\chi}_{v,I}:=\sqrt{(\hat{\eta}_{v,I}/t)^2+1/4}-1/2,
\end{equation}
which have the common  eigen-states $|\vec{j},\vec{\mathfrak{I}}\rangle$ with their eigen-values being given by the quantum numbers $j_e$ and $j_{v,I}$ as
\begin{equation}\label{chiaction}
\hat{\chi}_{e}|\vec{j},\vec{\mathfrak{I}}\rangle=j_e|\vec{j},\vec{\mathfrak{I}}\rangle,\quad \hat{\chi}_{v,I}|\vec{j},\vec{\mathfrak{I}}\rangle=j_{v,I}|\vec{j},\vec{\mathfrak{I}}\rangle.
\end{equation}
Hence, it is reasonable to suppose that  the operators   $\hat{\chi}_{e}$ and $\hat{\chi}_{v,I}$  are a set of fundamental operators in the Hilbert space ${\mathcal{H}}_\gamma^{\text{inv}}$ . Correspondingly, ${\chi}_e:=\sqrt{({\eta}_{e}/t)^2+1/4}-1/2$ and ${\chi}_{v,I}:=\sqrt{({\eta}_{v,I}/t)^2+1/4}-1/2$ are a set of classical fundamental  variables.
\item 
 In the second step, let us consider the Poisson brackets between the regularized variables ${e^{\pm\textbf{i}r(\eta_e)\varrho_e}}$, ${e^{\pm\textbf{i}r(\eta_e){\varrho}_{v,I}}}$ and ${\chi}_e$ and ${\chi}_{v,I}$ ,  which read
\begin{eqnarray}\label{Po3}
&&\{{e^{\pm\textbf{i}r(\eta_e)\varrho_e}},{\chi}_{e'}\} =\pm\frac{\textbf{i}}{\hbar}\delta_{e,e'}{e^{\pm\textbf{i}r(\eta_e)\varrho_e}},\\\nonumber
&&\{{e^{\pm\textbf{i}r(\eta_{v,I}){\varrho}_{v,I}}},{\chi}_{v',I'}\} =\pm\frac{\textbf{i}}{\hbar}\delta_{v,v'}\delta_{I,I'}{e^{\pm\textbf{i}r(\eta_{v,I}){\varrho}_{v,I}}},
\end{eqnarray}
where $r(\eta_e):=\frac{\sqrt{({\eta}_{e})^2+t^2/4}}{\eta_e}$ and $r(\eta_{v,I}):=\frac{\sqrt{({\eta}_{v,I})^2+t^2/4}}{\eta_{v,I}}$ are the regulators. 

The  quantum representation of the Poisson brackets \eqref{Po3} in the Hilbert space ${\mathcal{H}}_\gamma^{\text{inv}}$  could be given by considering the action of the corresponding operator commutators on the basis state$|\vec{j},\vec{\mathfrak{I}}\rangle$, which reads
\begin{eqnarray}\label{Poquan}
&&[\widehat{e^{\pm\textbf{i} r(\eta_e)\varrho_e}},\hat{\chi}_{e'}]|\vec{j},\vec{\mathfrak{I}}\rangle=\mp\delta_{e,e'}\widehat{e^{\pm\textbf{i}r(\eta_e)\varrho_e}}|\vec{j},\vec{\mathfrak{I}}\rangle,
\\\nonumber
&&[\widehat{e^{\pm\textbf{i} r(\eta_{v,I}){\varrho}_{v,I}}},\hat{\chi}_{v',I'}]|\vec{j},\vec{\mathfrak{I}}\rangle=\mp\delta_{v,v'}\delta_{I,I'}\widehat{e^{\pm\textbf{i}r(\eta_{v,I}){\varrho}_{v,I}}}|\vec{j},\vec{\mathfrak{I}}\rangle.
\end{eqnarray}
Indeed, the existence of this representation is determined by the solvability of Eqs.\eqref{Poquan}. By using Eqs.\eqref{chiaction}, one can solve Eqs.\eqref{Poquan} and it leads to the operators $\widehat{e^{\pm \textbf{i} r(\eta_e)\varrho_e}}$ and $\widehat{e^{\pm\textbf{i}r(\eta_{v,I}){\varrho}_{v,I}}}$ acting on the basis state $|\vec{j},\vec{\mathfrak{I}}\rangle$ as
\begin{eqnarray}\label{Poquan2}
\widehat{e^{\pm\textbf{i}r(\eta_e)\varrho_e}}|\vec{j},\vec{\mathfrak{I}}\rangle=\Xi(\vec{j}^{ j_e\pm1},\vec{\mathfrak{I}})|\vec{j}^{ j_e\pm1},\vec{\mathfrak{I}}\rangle, &&\quad 
\widehat{e^{\pm\textbf{i}r(\eta_{v,I}){\varrho}_{v,I}}}|\vec{j},\vec{\mathfrak{I}}\rangle=\Xi(\vec{j},\vec{\mathfrak{I}}^{j_{v,I}\pm1})|\vec{j},\vec{\mathfrak{I}}^{j_{v,I}\pm1}\rangle
\end{eqnarray}
 where
\begin{equation}
\vec{j}^{ j_e\pm1}=(...,j_{e'}\pm\delta_{e,e'},...)|_{e'\in E(\gamma)}
\end{equation}
and 
\begin{equation}
\vec{\mathfrak{I}}^{j_{v,I}\pm1}=(...,(j_{v',1}\pm\delta_{v,v'}\delta_{1,I},j_{v',2}\pm\delta_{v,v'}\delta_{2,I},j_{v',3}\pm\delta_{v,v'}\delta_{3,I}),...)|_{v'\in V(\gamma)};
\end{equation}
 Moreover, the functions $\Xi(\vec{j}^{ j_e\pm1},\vec{\mathfrak{I}})$ and $\Xi(\vec{j},\vec{\mathfrak{I}}^{j_{v,I}\pm1})$ are given by 
\begin{equation}\label{Xie}
\Xi(\vec{j}^{ j_e\pm1},\vec{\mathfrak{I}})=
\begin{cases}
1,& \text{if}\ (\vec{j}^{j_e\pm 1},\vec{\mathfrak{I}})\  \text{satisfy condition}\  \eqref{quanboundary}\ \text{for all}\  v\in  V(\gamma),\\
0,&\text{otherwise},
\end{cases}
\end{equation}
and 
\begin{equation}\label{Xiv}
\Xi(\vec{j},\vec{\mathfrak{I}}^{j_{v,I}\pm1})=
\begin{cases}
1,& \text{if}\ (\vec{j},\vec{\mathfrak{I}}^{j_{v,I}\pm1})\  \text{satisfy condition}\  \eqref{quanboundary}\ \text{for all}\  v\in  V(\gamma),\\
0,&\text{otherwise}.
\end{cases}
\end{equation}


\item 
Now, all of the quantum operators $\widehat{e^{\pm\textbf{i}r(\eta_e)\varrho_e}},\hat{\chi}_{e'},\widehat{e^{\pm\textbf{i}r(\eta_{v,I}){\varrho}_{v,I}}}$ and $\hat{\chi}_{v',I'}$  corresponding to the regularized variables ${e^{\pm\textbf{i}r(\eta_e)\varrho_e}},{\chi}_{e'},{e^{\pm\textbf{i}r(\eta_{v,I}){\varrho}_{v,I}}}$ and ${\chi}_{v',I'}$ have been established  in ${\mathcal{H}}_\gamma^{\text{inv}}$. It is easy to see that the regularized variables ${e^{\pm\textbf{i}r(\eta_e)\varrho_e}},{\chi}_{e'},{e^{\pm\textbf{i}r(\eta_{v,I}){\varrho}_{v,I}}}$ and ${\chi}_{v',I'}$ is a set of complete canonical coordinates for the  reduced twisted geometry space $\dot{H}^+_\gamma\cup \bar{H}^{'+}_\gamma$, while the corresponding quantum operators $\widehat{e^{\pm\textbf{i}r(\eta_e)\varrho_e}},\hat{\chi}_{e'},\widehat{e^{\pm\textbf{i}r(\eta_{v,I}){\varrho}_{v,I}}}$ and $\hat{\chi}_{v',I'}$ are able to generate the  Hilbert space  ${\mathcal{H}}^{\text{inv}}_\gamma$. Hence, one can claim that the Poisson algebra among ${e^{\pm\textbf{i}r(\eta_e)\varrho_e}},{\chi}_{e'},{e^{\pm\textbf{i}r(\eta_{v,I}){\varrho}_{v,I}}},{\chi}_{v',I'}$ is the fundamental algebra generating the Hilbert space  ${\mathcal{H}}_\gamma^{\text{inv}}$,
if the quantum algebras among the operators  $\widehat{e^{\pm\textbf{i}r(\eta_e)\varrho_e}},\hat{\chi}_{e'},\widehat{e^{\pm\textbf{i}r(\eta_{v,I}){\varrho}_{v,I}}},\hat{\chi}_{v',I'}$ give a faithful quantum representation of the Poisson algebra among ${e^{\pm\textbf{i}r(\eta_e)\varrho_e}},{\chi}_{e'},{e^{\pm\textbf{i}r(\eta_{v,I}){\varrho}_{v,I}}},{\chi}_{v',I'}$. According to above results, we still need to check if the quantum commutators among $\widehat{e^{\pm\textbf{i}r(\eta_e)\varrho_e}},\widehat{e^{\pm\textbf{i}r(\eta_{v,I}){\varrho}_{v,I}}}$ are isomorphic to the corresponding  Poisson brackets. The explicit analysis of this issue is shown in Appendix \ref{appPoisson}. The result shows that the quantum algebras among the operators  $\widehat{e^{\pm\textbf{i}r(\eta_e)\varrho_e}},\hat{\chi}_{e'},\widehat{e^{\pm\textbf{i}r(\eta_e){\varrho}_{v,I}}},\hat{\chi}_{v',I'}$ give a faithful quantum representation of the Poisson algebra among ${e^{\pm\textbf{i}r(\eta_e)\Xi^{\pm}_{e}(\vec{\chi})\varrho_e}}, {\chi}_{e'}, {e^{\pm\textbf{i}r(\eta_{v,I})\Xi^{\pm}_{v,I}(\vec{\chi}){\varrho}_{v,I}}}$ and ${\chi}_{v',I'}$ for some  specific regulators $\Xi^{\pm}_{e}(\vec{\chi})$ and $\Xi^{\pm}_{v,I}(\vec{\chi})$, with  ${e^{\pm\textbf{i}r(\eta_e)\Xi^{\pm}_{e}(\vec{\chi})\varrho_e}}$ and $ {e^{\pm\textbf{i}r(\eta_{v,I})\Xi^{\pm}_{v,I}(\vec{\chi}){\varrho}_{v,I}}}$  are distinguished with  ${e^{\pm\textbf{i}r(\eta_e)\varrho_e}}$ and ${e^{\pm\textbf{i}r(\eta_{v,I}){\varrho}_{v,I}}}$ up to $\mathcal{O}(t^n)$, with $t=\frac{\kappa\hbar}{a^2}$ and $n\gg1$ being a finite large real number. 
\end{enumerate}


Finally, we conclude that the fundamental algebra which generates the  Hilbert space ${\mathcal{H}}_\gamma^{\text{inv}}$  is given by the following Poisson algebra in the reduced twisted geometry space, 
\begin{eqnarray}\label{reducehoflux1}
&&\{\chi_{e'},{ \chi}_{e}\}=\{\chi_{v,I},{ \chi}_{v',I'}\}=\{\chi_{e},{ \chi}_{v,I}\}=0,
\\\nonumber
&&\{{e^{\pm\textbf{i}r(\eta_e)\Xi^{\pm}_{e}(\vec{\chi})\cdot\varrho_e}},{\chi}_{e'}\} =\pm\frac{\textbf{i}}{\hbar}\delta_{e,e'}\Xi^{\pm}_{e}(\vec{\chi})\cdot{e^{\pm\textbf{i}r(\eta_e)\Xi^{\pm}_{e}(\vec{\chi})\cdot\varrho_e}},
\\\nonumber
&&
 \{e^{\pm\textbf{i}r(\eta_{v,I})\Xi^{\pm}_{v,I}(\vec{\chi})\cdot{\varrho}_{v,I}},{\chi}_{v',I'}\} =\pm\frac{\textbf{i}}{\hbar}\delta_{v,v'}\delta_{I,I'}\Xi^{\pm}_{v,I}(\vec{\chi})\cdot e^{\pm\textbf{i}r(\eta_{v,I})\Xi^{\pm}_{v,I}(\vec{\chi})\cdot{\varrho}_{v,I}},
\end{eqnarray}
and 
\begin{eqnarray}\label{reducehoflux2}
\{{e^{\pm\textbf{i}r(\eta_e)\Xi^{\pm}_{e}(\vec{\chi})\cdot\varrho_e}},{e^{\pm\textbf{i}r(\eta_{e'})\Xi^{\pm}_{e'}(\vec{\chi})\cdot\varrho_{e'}}}\} &=&\begin{cases}
\text{Non-vanishing term},& \text{on specific boundary,}\\
0,&\text{otherwise,}
\end{cases}
\\\nonumber
 \{e^{\pm\textbf{i}r(\eta_{v,I})\Xi^{\pm}_{v,I}(\vec{\chi})\cdot{\varrho}_{v,I}},e^{\pm\textbf{i}r(\eta_{v',I'})\Xi^{\pm}_{v',I'}(\vec{\chi})\cdot{\varrho}_{v',I'}}\} &=&\begin{cases}
\text{Non-vanishing term},& \text{on specific boundary,}\\
0,&\text{otherwise,}
\end{cases}
\\\nonumber
 \{{e^{\pm\textbf{i}r(\eta_e)\Xi^{\pm}_{e}(\vec{\chi})\cdot\varrho_e}},e^{\pm\textbf{i}r(\eta_{v',I'})\Xi^{\pm}_{v',I'}(\vec{\chi})\cdot{\varrho}_{v',I'}}\} &=&\begin{cases}
\text{Non-vanishing term},& \text{on specific boundary,}\\
0,&\text{otherwise},
\end{cases}
\end{eqnarray}
where the ``\text{Non-vanishing term}'' is different for each Poisson bracket in Eq.\eqref{reducehoflux2}
, and its specific expression can be founded in the example considered in App.\ref{appPoisson}. Moreover, the corresponding quantum algebra for the Poisson algebra given by Eqs.\eqref{reducehoflux1} and \eqref{reducehoflux1}  is just the commutators among the operators $\widehat{e^{\pm\textbf{i}r(\eta_e)\varrho_e}},\hat{\chi}_{e'},\widehat{e^{\pm\textbf{i}r(\eta_{v,I}){\varrho}_{v,I}}}$ and $\hat{\chi}_{v',I'}$ acting in  ${\mathcal{H}}_\gamma^{\text{inv}}$ . Since ${\mathcal{H}}_\gamma^{\text{inv}}$  is the reduced space of the Hilbert 
 space $\mathcal{H}_\gamma$ generated by the quantum representation of holonomy-flux algebra, we can claim that the Poisson algebra given by Eqs.\eqref{reducehoflux1} and \eqref{reducehoflux1} is just the reduced holonomy-flux algebra with respect to Gauss constraint.
Furthermore, it is still necessary to clarify a key property of the fundamental operators $\widehat{e^{\pm\textbf{i}r(\eta_e)\varrho_e}},\hat{\chi}_{e'},\widehat{e^{\pm\textbf{i}r(\eta_{v,I}){\varrho}_{v,I}}}$ and $\hat{\chi}_{v',I'}$. Notice that $  \Theta_{\dot{H}_\gamma^+\cup\bar{H}^{'+}_\gamma}$ gives a degenerate symplectic 2-form on $\bar{H}^{'+}_\gamma$, and thus some operators in the set $\{\widehat{e^{\pm\textbf{i}r(\eta_e)\varrho_e}},\hat{\chi}_{e'},\widehat{e^{\pm\textbf{i}r(\eta_{v,I}){\varrho}_{v,I}}},\hat{\chi}_{v',I'}\}$ do not exist in some specific subspaces of the Hilbert space ${\mathcal{H}}_\gamma^{\text{inv}}$. For example,  the operators $(\hat{\chi}_{v',1},\widehat{e^{\pm\textbf{i}r(\eta_{v',1}){\varrho}_{v',1}}})$ do not exist in ${\mathcal{H}}_\gamma^{j_{e_1(v')=0}}\subset {\mathcal{H}}_\gamma^{\text{inv}}$, with  ${\mathcal{H}}_\gamma^{j_{e_1(v')=0}}$  spanned by the basis state $\{|\vec{j},\vec{\mathfrak{I}}\rangle|{j_{e_1(v')=0}}\}$ .

\subsection{Operator of the extrinsic curvature}\label{sec:402}

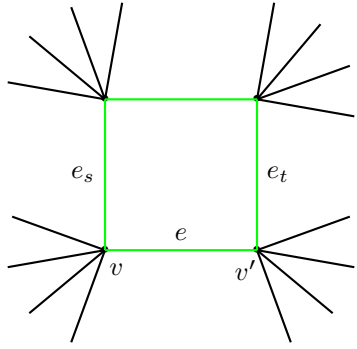
\begin{figure}[htb]
	\centering
	\begin{tikzpicture} [scale=1]

\coordinate  (A) at (0.5,1);

\coordinate  (B) at (2.5,3);

\coordinate  (C) at (0.5,3);

\coordinate  (D) at (2.5,1);

\node[scale=0.7] at (A) {$\bullet$} ;
\node[scale=0.7] at (B) {$\bullet$};
\node[scale=0.7] at (C) {$\bullet$};
\node[scale=0.7] at (D) {$\bullet$};
\draw (A) ++ (300:0.3) node {$v$};
\draw (D) ++ (240:0.3) node {$v'$};

\draw[thick] (B)  --  node[midway,sloped]{} ++ (50:1.3) node[right] {};
\draw[thick] (B)  to  node[midway,sloped]{} ++ (-10:1.3) node[right]{};
\draw[thick] (B)  --  node[midway,sloped]{} ++ (80:1.3) node[right] {};
\draw[thick] (B)  to  node[midway,sloped]{} ++ (20:1.3) node[right]{};

\draw[thick] (A)  to  node[midway,sloped]{} ++ (160:1.3) node[right]{};
\draw[thick] (A)  to  node[midway,sloped]{} ++ (190:1.3) node[left]{};
\draw[thick] (A)  to  node[midway,sloped]{} ++ (250:1.3) node[right]{};
\draw[thick] (A)  to  node[midway,sloped]{} ++ (220:1.3) node[left]{};

\draw[thick] (C)  to  node[midway,sloped]{} ++ (80:1.3) node[right]{};
\draw[thick] (C)  to  node[midway,sloped]{} ++ (110:1.3) node[left]{};
\draw[thick] (C)  to  node[midway,sloped]{} ++ (140:1.3) node[right]{};
\draw[thick] (C)  to  node[midway,sloped]{} ++ (170:1.3) node[left]{};

\draw[thick] (D)  to  node[midway,sloped]{} ++ (20:1.3) node[right]{};
\draw[thick] (D)  to  node[midway,sloped]{} ++ (-10:1.3) node[left]{};
\draw[thick] (D)  to  node[midway,sloped]{} ++ (-40:1.3) node[right]{};
\draw[thick] (D)  to  node[midway,sloped]{} ++ (-70:1.3) node[left]{};

\draw[thick,green] (A) -- node[midway,sloped]{} node[above,black] {$e$} (D);
\draw[thick,green] (B) -- node[midway,sloped]{} node[above,black] {} (C);
\draw[thick,green] (D) -- node[midway,sloped]{} node[right,black] {$e_t$} (B);
\draw[thick,green] (C) -- node[midway,sloped]{} node[left,black] {$e_s$} (A);

\end{tikzpicture}
\caption{The illustration of the minimal loop $\square_{e}\ni e_s, e_t$.}
\label{figresquarevv'}
\end{figure}

It is ready to consider the quantization of the densitized extrinsic curvatures  $\mathcal{K}_e^{\ e}$, $\mathcal{K}_e^{\ e'}$ and $\mathcal{K}_e^{\ \tilde{e}'}$.
To define the non-polynomial functions of flux operators, let us introduce the notations
\begin{equation}\label{Yinverse}
\hat{Y}^{-1}:=\sum_{Y\in\mathcal{E}\setminus 0}{Y}^{-1}|Y\rangle\langle Y|
\end{equation}
and
\begin{equation}
\Big(\sqrt{\hat{Y}}\Big)^{-1}:=\sum_{Y\in\mathcal{E}\setminus 0}\sqrt{Y}^{-1}|Y\rangle\langle Y|,
\end{equation}
where $\mathcal{E}$ is the eigen-spectrum of $\hat{Y}$ and $|Y\rangle$ is the eigen-state which corresponding to the eigenvalue $Y$ of $\hat{Y}$. Then, $\Big(\sqrt{{p}_e^i{p}_{e,i}+\frac{t^2}{4}}\Big)^{-1}$  can be promoted as the operator $\Big(\sqrt{\hat{p}_e^i\hat{p}_{e,i}+\frac{t^2}{4}}\Big)^{-1}$.
Now,  we can define the operator $\hat{ \mathcal{K}}^i_{e,s(e)}$ and $\hat{ \mathcal{K}}^i_{e,t(e)}$ which correspond to the regularized extrinsic curvature 1-form given in Eq.\eqref{regu1form}, which reads
\begin{equation}\label{K1formquan1}
\hat{ \mathcal{K}}^j_{e,s(e)}:=\frac{\hat{\Omega}_e\hat{p}^j_{e}\Big(\sqrt{\hat{p}_e^i\hat{p}_{e,i}+\frac{t^2}{4}}\Big)^{-1}+\hat{p}^j_{e}\Big(\sqrt{\hat{p}_e^i\hat{p}_{e,i}+\frac{t^2}{4}}\Big)^{-1}\hat{\Omega}_e}{2\beta}
\end{equation}
and 
\begin{equation}\label{K1formquan2}
\hat{ \mathcal{K}}^j_{e,t(e)}:=-\frac{\hat{\Omega}_e\hat{\tilde{p}}^j_{e}\Big(\sqrt{\hat{p}_e^i\hat{p}_{e,i}+\frac{t^2}{4}}\Big)^{-1}+\hat{\tilde{p}}^j_{e}\Big(\sqrt{\hat{p}_e^i\hat{p}_{e,i}+\frac{t^2}{4}}\Big)^{-1}\hat{\Omega}_e}{2\beta},
\end{equation}
where we defined
\begin{eqnarray}
\hat{\Omega}_e&:=&\frac{(\widehat{e^{\textbf{i}r(\eta_e)\varrho_e}}-\widehat{e^{-\textbf{i}r(\eta_e)\varrho_e}})}{2\textbf{i}}.
\end{eqnarray}
In addition, the operators $\hat{\mathcal{K}}_e^{\ e}$, $\hat{\mathcal{K}}_e^{\ e_s}$ and $\hat{\mathcal{K}}_e^{\ e_t}$ for the densitized extrinsic curvatures \eqref{Kdef1},\eqref{Kdef2} and \eqref{Kdef3} on graph $\gamma$  can be defined accordingly.  As shown in Fig.\ref{figresquarevv'},  one needs to fix a choice of minimal loops for the whole cubic graph, which ensures that the minimal loop $\square_e\ni e$ also satisfies  $\square_{e}\ni e_s, e_t$.  Then, the definition of the operators $\hat{\mathcal{K}}_e^{\ e}$, $\hat{\mathcal{K}}_e^{\ e_s}$ and $\hat{\mathcal{K}}_e^{\ e_t}$ are associated to $\square_e$. Without loss of generality, we can re-orient the edges to ensure $s(e)=s(e_s)=v$, $t(e)=v'$, $s(e_v)=v'$ and set $e=e_1(v)=e_1(v')$, $e_s=e_2(v)$, $e_t=e_2(v')$. Further, the operators $\hat{\mathcal{K}}_e^{\ e}$, $\hat{\mathcal{K}}_e^{\ e_s}$ and $\hat{\mathcal{K}}_e^{\ e_t}$ can be defined as
\begin{eqnarray}\label{K11}
 \hat{\mathcal{K}}_e^{\ e}&:=&\frac{\hat{\Omega}_e\hat{p}^j_{e}\hat{p}_{e,j}\Big(\sqrt{\hat{p}_e^i\hat{p}_{e,i}+\frac{t^2}{4}}\Big)^{-1}+\hat{p}^j_{e}\hat{p}_{e,j}\Big(\sqrt{\hat{p}_e^i\hat{p}_{e,i}+\frac{t^2}{4}}\Big)^{-1}\hat{\Omega}_e}{2\beta},
\end{eqnarray}
\begin{eqnarray}\label{K12}
 \hat{\mathcal{K}}_e^{\ e_s}
&:=&\frac{\hat{\Omega}_e\hat{p}_{e,j}\hat{p}^j_{e_s}\Big(\sqrt{\hat{p}_e^k\hat{p}_{e,k}+\frac{t^2}{4}}\Big)^{-1}+\hat{p}_{e,j}\hat{p}^j_{e_s}\Big(\sqrt{\hat{p}_e^k\hat{p}_{e,k}+\frac{t^2}{4}}\Big)^{-1}\hat{\Omega}_e}{2\beta},   
\end{eqnarray}
and 
\begin{eqnarray}\label{K21}
  \hat{\mathcal{K}}_e^{\ e_t}&:=&\frac{\hat{\Omega}_e\hat{\tilde{p}}_{e,j}\hat{p}^j_{e_t}\Big(\sqrt{\hat{p}_e^k\hat{p}_{e,k}+\frac{t^2}{4}}\Big)^{-1}+\hat{\tilde{p}}_{e,j}\hat{p}^j_{e_t}\Big(\sqrt{\hat{p}_e^k\hat{p}_{e,k}+\frac{t^2}{4}}\Big)^{-1}\hat{\Omega}_e}{2\beta}.
\end{eqnarray}

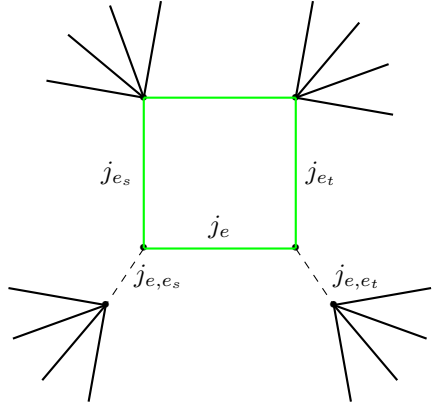
\begin{figure}[htb]
	\centering
	\begin{tikzpicture} [scale=1]

\coordinate  (A) at (0.5,1);

\coordinate  (AA) at (0,0.25);
\coordinate  (B) at (2.5,3);

\coordinate  (C) at (0.5,3);

\coordinate  (D) at (2.5,1);
\coordinate  (DD) at (3,0.25);

\node[scale=0.7] at (A) {$\bullet$} ;
\node[scale=0.7] at (AA) {$\bullet$} ;
\node[scale=0.7] at (B) {$\bullet$};
\node[scale=0.7] at (C) {$\bullet$};
\node[scale=0.7] at (D) {$\bullet$};
\node[scale=0.7] at (DD) {$\bullet$};

\draw[thick] (B)  --  node[midway,sloped]{} ++ (50:1.3) node[right] {};
\draw[thick] (B)  to  node[midway,sloped]{} ++ (-10:1.3) node[right]{};
\draw[thick] (B)  --  node[midway,sloped]{} ++ (80:1.3) node[right] {};
\draw[thick] (B)  to  node[midway,sloped]{} ++ (20:1.3) node[right]{};

\draw[thick] (AA)  to  node[midway,sloped]{} ++ (170:1.3) node[right]{};
\draw[thick] (AA)  to  node[midway,sloped]{} ++ (200:1.3) node[left]{};
\draw[thick] (AA)  to  node[midway,sloped]{} ++ (260:1.3) node[right]{};
\draw[thick] (AA)  to  node[midway,sloped]{} ++ (230:1.3) node[left]{};

\draw[thick] (C)  to  node[midway,sloped]{} ++ (80:1.3) node[right]{};
\draw[thick] (C)  to  node[midway,sloped]{} ++ (110:1.3) node[left]{};
\draw[thick] (C)  to  node[midway,sloped]{} ++ (140:1.3) node[right]{};
\draw[thick] (C)  to  node[midway,sloped]{} ++ (170:1.3) node[left]{};

\draw[thick] (DD)  to  node[midway,sloped]{} ++ (10:1.3) node[right]{};
\draw[thick] (DD)  to  node[midway,sloped]{} ++ (-20:1.3) node[left]{};
\draw[thick] (DD)  to  node[midway,sloped]{} ++ (-50:1.3) node[right]{};
\draw[thick] (DD)  to  node[midway,sloped]{} ++ (-80:1.3) node[left]{};

\draw[thick,green] (A) -- node[midway,sloped]{} node[above,black] {$j_{e}$} (D);
\draw[thick,green] (B) -- node[midway,sloped]{} node[above,black] {} (C);
\draw[thick,green] (D) -- node[midway,sloped]{} node[right,black] {$j_{e_t}$} (B);
\draw[thick,green] (C) -- node[midway,sloped]{} node[left,black] {$j_{e_s}$} (A);
\draw[dashed] (A) -- node[midway,sloped]{} node[right,black] {$j_{e,e_s}$} (AA);
\draw[dashed] (D) -- node[midway,sloped]{} node[right,black] {\ $j_{e,e_t}$} (DD);

\end{tikzpicture}
\caption{The recoupling of spin $j_e$ and $j_{e_s}$ gives spin $j_{e,e_s}$ which labels an internal edge associated to vertex $v$, and the recoupling of spin $j_e$ and $j_{e_t}$ gives spin $j_{e,e_t}$ which labels an internal edge associated to vertex $v'$.}
\label{figresquarevv'j}
\end{figure}

The spin-network basis $|\vec{j},\vec{\mathfrak{I}}\rangle$ on the cubic graph can be adapted to the notations for the partial graph related to $\square_e$, which gives $j_e=j_{e_1(v)}=j_{e_1(v')}$, $j_{e_s}=j_{e_2(v)}$, $j_{e_t}=j_{e_2(v')}$ , $j_{e,e_s}=j_{v,1}$ and $j_{e,e_t}=j_{v',1}$, see the illustration in Fig.\ref{figresquarevv'j}.
Then, the action of the operators $\hat{\mathcal{K}}_e^{\ e}$, $\hat{\mathcal{K}}_e^{\ e_s}$ and $\hat{\mathcal{K}}_e^{\ e_t}$  on the state $|\vec{j},\vec{\mathfrak{I}}\rangle$ can be given by their component operators
\begin{equation}
\hat{p}^j_{e}\hat{p}_{e,j}|\vec{j},\vec{\mathfrak{I}}\rangle=t^2j_e(j_e+1)|\vec{j},\vec{\mathfrak{I}}\rangle,   
\end{equation}
\begin{equation}
 \hat{p}_{e,j}\hat{p}^j_{e_s}|\vec{j},\vec{\mathfrak{I}}\rangle=\frac{t^2}{2}(j_{e,e_s}(j_{e,e_s}+1)-j_e(j_e+1)-j_{e_s}(j_{e_s}+1))|\vec{j},\vec{\mathfrak{I}}\rangle,   
\end{equation}
\begin{equation}
 \hat{\tilde{p}}_{e,j}\hat{p}^j_{e_t}|\vec{j},\vec{\mathfrak{I}}\rangle=\frac{t^2}{2}(j_{e,e_t}(j_{e,e_t}+1)-j_e(j_e+1)-j_{e_t}(j_{e_t}+1))|\vec{j},\vec{\mathfrak{I}}\rangle,   
\end{equation}
and 
\begin{equation}
\hat{\Omega}_e|\vec{j},\vec{\mathfrak{I}}\rangle=\frac{1}{2\textbf{i}}\left(\Xi(\vec{j}^{ j_e\pm1},\vec{\mathfrak{I}})|\vec{j}^{j_e+1},\vec{\mathfrak{I}}\rangle+\Xi(\vec{j},\vec{\mathfrak{I}}^{j_{v,I}\pm1})|\vec{j}^{j_e-1},\vec{\mathfrak{I}}\rangle\right),   
\end{equation}
where  $\Xi(\vec{j}^{ j_e\pm1},\vec{\mathfrak{I}})$ and $\Xi(\vec{j},\vec{\mathfrak{I}}^{j_{v,I}\pm1})$ are given by Eqs.\eqref{Xie} and \eqref{Xiv}. 

Notice that the eigen spectrum are clear for the operators $\hat{p}^j_{e}\hat{p}_{e,j}$, $ \hat{p}_{e,j}\hat{p}^j_{e_s}$ and  $ \hat{\tilde{p}}_{e,j}\hat{p}^j_{e_t}$. It is also worth to analyze the spectrum of 
the operator $\hat{\Omega}_e$. For fixed  $\vec{j}_{\setminus e}:=\vec{j}\setminus \{j_e\}$ and $\vec{\mathfrak{I}}$, one can
consider the eigen equation 
\begin{equation}
\hat{\Omega}_e|\lambda_e,\vec{j}_{\setminus e},\vec{\mathfrak{I}}\rangle=\lambda_e|\lambda_e,\vec{j}_{\setminus e},\vec{\mathfrak{I}}\rangle,
\end{equation}
where   $$|\lambda_e,\vec{j}_{\setminus e},\vec{\mathfrak{I}}\rangle=\sum_{j_e=j_e^{\text{min}}}^{j_e^{\text{max}}}c_{\lambda_e}(j_e)|\vec{j},\vec{\mathfrak{I}}\rangle $$ is the eigen state of $\hat{\Omega}_e$ with eigen value $\lambda_e$, and  $j_e^{\text{min}}=\min\{|j_e-j_{e,e_s}|,|j_e-j_{e,e_t}|\}$,  $j_e^{\text{max}}=\max\{|j_e+j_{e,e_s}|,|j_e+j_{e,e_t}|\}$. This equation leads to the difference equation 
\begin{equation}
c_{j_e-1}-c_{j_e+1}=2\textbf{i}\lambda_e  c_{j_e}, \  \text{for}\  j_e<j_e^{\text{max}} \ \text{and}\  j_e>j_e^{\text{min}},
\end{equation}
and 
\begin{equation}
c_{j^{\text{max}}_e-1}=2\textbf{i}\lambda_e  c_{j^{\text{max}}_e},\quad -c_{j^{\text{min}}_e+1}=2\textbf{i}\lambda_e  c_{j^{\text{min}}_e}.
\end{equation}
 By solving the difference equation, one can get the eigenvalue
\begin{equation}\label{eigenlambda}
\lambda_e =\frac{e^{\textbf{i}\theta}+e^{-\textbf{i}\theta}}{2}
\end{equation}
where $\theta$ is determined by
\begin{equation}\label{eigentheta}
 0<\theta=\frac{k \pi}{j_e^{\text{max}}-j_e^{\text{min}}+2}\leq\pi, k=1,2,3,...
\end{equation}
Correspondingly, the  expansion  coefficients $c_{j_e}$ of the  eigenstate $|\lambda_e,\vec{j}_{\setminus e},\vec{\mathfrak{I}}\rangle$is given by
\begin{equation}
c_{\lambda_e}(j_e)=(-\textbf{i})^{j_e}C_1\exp(\textbf{i}j_e\theta)+(-\textbf{i})^{j_e}C_2\exp(-\textbf{i}j_e\theta),
\end{equation}
where 
\begin{eqnarray}
C_1/C_2=-\exp(-2\textbf{i}(j^{\max}_e+1)\theta)=-\exp(-2\textbf{i}(j_e^{\min}-1)\theta).
\end{eqnarray}
The Eqs.\eqref{eigenlambda} and \eqref{eigentheta} show that $\lambda_e$ takes discrete value in the range $-1\leq \lambda_e< 1$ for fixed   $\vec{j}_{\setminus e}:=\vec{j}\setminus \{j_e\}$ and $\vec{\mathfrak{I}}$. Nevertheless, one should notice that the ranges of $\vec{j}_{\setminus e}:=\vec{j}\setminus \{j_e\}$ and $\vec{\mathfrak{I}}$ ensure $j_e^{\text{max}}-j_e^{\text{min}}\in\frac{\mathbb{N}_+}{2}$ taking values in the range $0\leq j_e^{\text{max}}-j_e^{\text{min}}\leq +\infty$. Thus, it is to see that  $\lambda_e$ takes continuum value in the range $-1\leq \lambda_e< 1$ in the gauge invariant Hilbert space $\mathcal{H}^{\text{inv}}_{\gamma}$ on cubic graph $\gamma$.
Finally, one can conclude that the operator $\hat{\Omega}_e$ has the continuum spectrum
\begin{equation}
\text{Spec}(\hat{\Omega}_e)=\{\lambda|-1\leq\lambda<1\}.
\end{equation}

It is worth to compare the extrinsic curvature operators \eqref{K1formquan1} and \eqref{K1formquan2}  to the operator \eqref{Kexist}, which represent the regularized extrinsic curvature 1-form in LQG. One should notice that the operator \eqref{Kexist}
 contains the commutators among the holonomy operator, volume operator and the Euclidean part of the scalar constraint operator, which leads that its action on the spin-network state is rather complicated. Nevertheless, the new extrinsic curvature operators \eqref{K1formquan1} and \eqref{K1formquan2}  established in this article have clearer and simpler actions on the spin-network states. It is expected that the obstacles caused by the  complication of extrinsic curvature operator in previous studies can be avoided by using our new   extrinsic curvature operators.

\section{Conclusion and Discussion}\label{sec5}
With the gauge invariant holonomy-flux phase space being parametrized by the reduced twisted geometry variables, the fundamental algebra of the reduced twisted geometry is specified in this article. More explicitly, this specification ensures that  the  quantum representation of the fundamental algebra in the reduced twisted geometry gives the gauge invariant Hilbert space in LQG. Correspondingly, the fundamental operators which represented the reduced twisted geometry is established. Based on these fundamental operators,  a new type of extrinsic curvature operator is constructed in LQG.

Additionally, there are several points worth to be discussed. First, the  reduced twisted geometry variables forms a simple Poisson algebra in the gauge invariant holonomy-flux phase space. In this article, we introduce a regularization of  the  reduced twisted geometry variables to construct  the fundamental algebra whose quantum representation generates the gauge invariant Hilbert space in LQG. Nevertheless, one may consider the quantum representation of the Poisson algebra among the reduced twisted geometry variables directly, e.g. the Bohr-Sommerfeld quantization as shown in \cite{PhysRevLett.107.011301,Bianchi:2012wb}. 
Second, the spatial intrinsic geometry sector in the reduced twisted geometry is composed by the polyhedra geometries associated to each vertex of the graph. In fact,  the quantization of the polyhedron has been studied based on the geometric quantization methods in several previous works \cite{PhysRevD.83.044035,Long:2020agv,Freidel:2010xna,long2019coherent,Conrady:2009px,Long:2020euh}, in which the quantum polyhedra is described by the intertwiners. It is reasonable to extend this quantization methods to the full twisted geometry, which is also expected to give the gauge invariant Hilbert space in LQG.
Third, since the extrinsic curvature operator constructed in this article is much simpler than the previous one, it is reasonable to apply it to define a new scalar constraint operator.  Moreover, one can also apply it to define the ADM energy operator and surface gravity operator, which are key observables in the  loop quantum black hole.  
Fourth, the twisted geometry parametrization for the holonomy-flux phase space has been extended to the higher dimension LQG based on the gauge group $SO(D+1)$ \cite{Bodendorfer:Ha,PhysRevD.103.086016,long2020operators,Long:2020agv,Long:2022thb,Long:2021lmd,Long:2022cex}. It is also interesting to check whether the simple algebras of the reduced twisted geometric variables can be extended to the $SO(D+1)$ holonomy-flux phase space \cite{Long:2023ivt}.
\section*{Acknowledgments}
This work is supported by the project funded by  the National Natural Science Foundation of China (NSFC) with Grants No. 12405062 and No.12275022. G. L. is supported by the Fundamental Research Funds for the Central Universities with Grants No.21624340, and the Science and Technology Planning Project of Guangzhou with Grants No. 2024A04J4030.  H. L. is supported by research grants provided by the Blaumann Foundation. 

\bibliographystyle{unsrt}

\bibliography{ref}

\appendix

\section{Regularization and quantization of the angle variable $\varrho_e$}\label{appPoisson}

It is necessary to ask whether the quantum algebras among the operators  $\widehat{e^{\pm\textbf{i}r(\eta_e)\varrho_e}},\hat{\chi}_{e'},\widehat{e^{\pm\textbf{i}r(\eta_e){\varrho}_{v,I}}},\hat{\chi}_{v',I'}$ give a faithful quantum representation of the Poisson algebra among ${e^{\pm\textbf{i}r(\eta_e)\varrho_e}},{\chi}_{e'},{e^{\pm\textbf{i}r(\eta_e){\varrho}_{v,I}}},{\chi}_{v',I'}$. Indeed, the answer is not. This can be seen by comparing the commutators among $\widehat{e^{\pm\textbf{i}r(\eta_e)\varrho_e}},\widehat{e^{\pm\textbf{i}r(\eta_e){\varrho}_{v,I}}}$ and the corresponding vanishing Poisson brackets among ${e^{\pm\textbf{i}r(\eta_e)\varrho_e}},{e^{\pm\textbf{i}r(\eta_e){\varrho}_{v,I}}}$. It is directly to see that
\begin{eqnarray}\label{noncom01}
&&[\widehat{e^{\textbf{i}r(\eta_{e})\varrho_{e}}},\widehat{e^{\textbf{i}r(\eta_{e'})\varrho_{e'}}}]|\vec{j},\vec{\mathfrak{I}}\rangle=[\widehat{e^{\textbf{i}r(\eta_{e}){\varrho}_{e}}},\widehat{e^{\textbf{i}r(\eta_{v,I}){\varrho}_{v,I}}}]|\vec{j},\vec{\mathfrak{I}}\rangle\\\nonumber
&=&[\widehat{e^{\textbf{i}r(\eta_{v,I}){\varrho}_{v,I}}},\widehat{e^{\textbf{i}r(\eta_{v',I'}){\varrho}_{v',I'}}}]|\vec{j},\vec{\mathfrak{I}}\rangle=0
\end{eqnarray}
for
\begin{eqnarray}\label{quanboundaryjjj}
&& |j_{e_1(v)}-j_{e_2(v)}|< { j}_{v,1}<j_{e_1(v)}+j_{e_2(v)},\\\nonumber
 &&\ 
  \  |j_{e_3(v)}-j_{e_4(v)}|<{ j}_{v,2}< j_{e_3(v)}+j_{e_4(v)},\quad |j_{e_5(v)}-j_{e_6(v)}|< {j}_{v,3}<j_{e_5(v)}+j_{e_6(v)},\\\nonumber
  &&\  \  |{ j}_{v,1}-{ j}_{v,2}|< { j}_{v,3}<{j}_{v,1}+{ j}_{v,2},\quad j_e>0,\ \forall e\in E(\gamma),\ v\in V(\gamma).
\end{eqnarray}
However, if one consider the commutators among $\widehat{e^{\pm\textbf{i}r(\eta_e)\varrho_e}},\widehat{e^{\pm\textbf{i}r(\eta_e){\varrho}_{v,I}}}$ acting on the states $|\vec{j},\vec{\mathfrak{I}}\rangle$ labelled by  the spins which do not satisfy the conditions in Eq.  \eqref{quanboundaryjjj}, one can get some non-vanished commutators; for instance, one has the  commutator
\begin{eqnarray}\label{noncom1}
&&[\widehat{e^{\textbf{i}r(\eta_{e_1})\varrho_{e_1}}},\widehat{e^{\textbf{i}r(\eta_{e_2})\varrho_{e_2}}}]|\vec{j},\vec{\mathfrak{I}}\rangle=\widehat{e^{\textbf{i}r(\eta_{e_1})\varrho_{e_1}}}\widehat{e^{\textbf{i}r(\eta_{e_2})\varrho_{e_2}}}|\vec{j},\vec{\mathfrak{I}}\rangle=\left.|\vec{j},\vec{\mathfrak{I}}\rangle\right|_{j_{e_1(v)}\to j_{e_1(v)}+1, j_{e_2(v)}\to j_{e_2(v)}+1}
\end{eqnarray}
for $j_{e_1(v)}=j_{e_2(v)}+j_{v,1}$, and 
\begin{eqnarray}\label{noncom2}
&&[\widehat{e^{\textbf{i}r(\eta_{v,1}){\varrho}_{v,1}}},\widehat{e^{\textbf{i}r(\eta_{v,2}){\varrho}_{v,2}}}]|\vec{j},\vec{\mathfrak{I}}\rangle=\widehat{e^{\textbf{i}r(\eta_{v,1}){\varrho}_{v,1}}}\widehat{e^{\textbf{i}r(\eta_{v,2}){\varrho}_{v,2}}}|\vec{j},\vec{\mathfrak{I}}\rangle=\left.|\vec{j},\vec{\mathfrak{I}}\rangle\right|_{j_{v,1}\to j_{v,1}+1, j_{v,2}\to j_{v,2}+1}
\end{eqnarray}
for $j_{v,1}=j_{v,2}+j_{v,3}$. It is easy to see that such non-vanished commutators are not the quantum representation of the corresponding Poisson brackets.

To find the correct Poisson algebras whose quantum representation is given by the quantum commutators among the operators  $\widehat{e^{\pm\textbf{i}r(\eta_e)\varrho_e}},\hat{\chi}_{e'},\widehat{e^{\pm\textbf{i}r(\eta_e){\varrho}_{v,I}}},\hat{\chi}_{v',I'}$ , it is necessary to introduce further regularization to ${e^{\pm\textbf{i}r(\eta_e)\varrho_e}}$ and $ {e^{\pm\textbf{i}r(\eta_{v,I}){\varrho}_{v,I}}}$.  Consider the regulators $\Xi^{\pm}_{e}(\vec{\chi})$ and $\Xi^{\pm}_{v,I}(\vec{\chi})$ , and then regularize ${e^{\pm\textbf{i}r(\eta_e)\varrho_e}}$ and $ {e^{\pm\textbf{i}r(\eta_{v,I}){\varrho}_{v,I}}}$ as ${e^{\pm\textbf{i}r(\eta_e)\Xi^{\pm}_{e}(\vec{\chi})\varrho_e}}$ and $ {e^{\pm\textbf{i}r(\eta_{v,I})\Xi^{\pm}_{v,I}(\vec{\chi}){\varrho}_{v,I}}}$ respectively.  The regulators $\Xi^{\pm}_{e}(\vec{\chi})$ and $\Xi^{\pm}_{v,I}(\vec{\chi})$, as functions of $\vec{\chi}=(...,\chi_{e},...,(\chi_{v,1},\chi_{v,2},\chi_{v,3},),...)$, are constrained by requiring that the quantum algebras among the operators $\widehat{e^{\pm\textbf{i}r(\eta_e)\varrho_e}},\hat{\chi}_{e'},\widehat{e^{\pm\textbf{i}r(\eta_e){\varrho}_{v,I}}},\hat{\chi}_{v',I'}$ give a faithful quantum representation of the Poisson algebra among ${e^{\pm\textbf{i}r(\eta_e)\Xi^{\pm}_{e}(\vec{\chi})\varrho_e}}, {\chi}_{e'}, {e^{\pm\textbf{i}r(\eta_{v,I})\Xi^{\pm}_{v,I}(\vec{\chi}){\varrho}_{v,I}}}$ and ${\chi}_{v',I'}$. 
To clarify the property of the regulators $\Xi^{\pm}_{e}(\vec{\chi})$ and $\Xi^{\pm}_{v,I}(\vec{\chi})$, let us consider their quantization
\begin{equation}
 \widehat{\Xi^{\pm}_{e}}(\vec{\chi}):=\sum_{\vec{j},\vec{\mathfrak{I}}}\Xi^{\pm}_{e}(\vec{\chi}_{\vec{j},\vec{\mathfrak{I}}})|\vec{j},\vec{\mathfrak{I}}\rangle\langle \vec{j},\vec{\mathfrak{I}}|
\end{equation}
\begin{equation}
\widehat{\Xi^{\pm}_{v,I}}(\vec{\chi}):=\sum_{\vec{j},\vec{\mathfrak{I}}}\Xi^{\pm}_{v,I}(\vec{\chi}_{\vec{j},\vec{\mathfrak{I}}})|\vec{j},\vec{\mathfrak{I}}\rangle\langle \vec{j},\vec{\mathfrak{I}}|,
\end{equation}
where $\vec{\chi}_{\vec{j},\vec{\mathfrak{I}}}:=(...,j_{e},...,(j_{v,1},j_{v,2},j_{v,3},),...)$. Then,  by requiring the isomorphic between the quantum commutator \eqref{Poquan} and the quantum representation of the Poisson brackets
\begin{equation}\label{Po555666}
\{{e^{\pm\textbf{i}r(\eta_e)\Xi^{\pm}_{e}(\vec{\chi})\cdot\varrho_e}},{\chi}_{e'}\} =\pm\frac{\textbf{i}}{\hbar}\delta_{e,e'}\Xi^{\pm}_{e}(\vec{\chi})\cdot{e^{\pm\textbf{i}r(\eta_e)\Xi^{\pm}_{e}(\vec{\chi})\cdot\varrho_e}},
\end{equation}
and
\begin{equation}\label{Po555666777}
 \{e^{\pm\textbf{i}r(\eta_{v,I})\Xi^{\pm}_{v,I}(\vec{\chi})\cdot{\varrho}_{v,I}},{\chi}_{v',I'}\} =\pm\frac{\textbf{i}}{\hbar}\delta_{v,v'}\delta_{I,I'}\Xi^{\pm}_{v,I}(\vec{\chi})\cdot e^{\pm\textbf{i}r(\eta_{v,I})\Xi^{\pm}_{v,I}(\vec{\chi})\cdot{\varrho}_{v,I}},
\end{equation}
 one can  conclude that the functions $\Xi^{\pm}_{e}(\vec{\chi})$ and $\Xi^{\pm}_{v,I}(\vec{\chi})$ must satisfy
\begin{equation}\label{reg1}
 \Xi^{\pm}_{e}(\vec{\chi})=\Xi^{\pm}_{v,I}(\vec{\chi})=1,\ \text{for}\  \chi_e\in \frac{\mathbb{N}}{2}, \ \text{and} \  \chi_{v,I}\in \frac{\mathbb{N}}{2}.
\end{equation}
Further,  let us consider  the Poisson brackets
 \begin{eqnarray}\label{Poxixi}
&&
\{{e^{\textbf{i}r(\eta_{e_1})\Xi^{+}_{e_1}(\vec{\chi})\cdot\varrho_{e_1}}},{e^{\textbf{i}r(\eta_{e_2})\Xi^{+}_{e_2}(\vec{\chi})\cdot\varrho_{e_2}}}\} \\\nonumber
&=&(-r(\eta_{e_2})r(\eta_{e_1})\varrho_{e_2}\Xi^{+}_{e_1}(\vec{\chi})\cdot\{\varrho_{e_1},\Xi^{+}_{e_2}(\vec{\chi})\}+r(\eta_{e_2})r(\eta_{e_1}) \varrho_{e_1} \Xi^{+}_{e_2}(\vec{\chi})\cdot\{\varrho_{e_2},\Xi^{+}_{e_1}(\vec{\chi})\})\\\nonumber
&&\cdot{e^{\textbf{i}r(\eta_{e_1})\Xi^{+}_{e_1}(\vec{\chi})\cdot\varrho_{e_1}}}{e^{\textbf{i}r(\eta_{e_2})\Xi^{+}_{e_2}(\vec{\chi})\cdot\varrho_{e_2}}}.
\end{eqnarray}
The isomorphism between the quantum representation of the Poisson bracket  \eqref{Poxixi} and the quantum commutator \eqref{noncom1}
can be established in the region defined by \eqref{quanboundary}. Specifically, by comparing  the quantum representation of Eq.  \eqref{Poxixi} to Eq.\eqref{noncom1}, it is easy see that  they can always be isomorphic to each other by adjusting the value of $\{\varrho_e,\Xi^{\pm}_{e'}(\vec{\chi})\}$. For instance, one can adjust the value of $\{\varrho_{e_1},\Xi^{+}_{e_2}(\vec{\chi})\}$ and  $\{\varrho_{e_2},\Xi^{+}_{e_1}(\vec{\chi})\}$ to ensure 
 \begin{eqnarray}\label{PoKK1}
&&
\{{e^{\textbf{i}r(\eta_{e_1})\Xi^{+}_{e_1}(\vec{\chi})\cdot\varrho_{e_1}}},{e^{\textbf{i}r(\eta_{e_2})\Xi^{+}_{e_2}(\vec{\chi})\cdot\varrho_{e_2}}}\} \\\nonumber
&=&\begin{cases}
\frac{2}{\textbf{i}\hbar}{e^{\textbf{i}r(\eta_{e_1})\Xi^{+}_{e_1}(\vec{\chi})\cdot\varrho_{e_1}}}{e^{\textbf{i}r(\eta_{e_2})\Xi^{+}_{e_2}(\vec{\chi})\cdot\varrho_{e_2}}},& \text{if}\ \chi_{e_1(v)}=\chi_{e_2(v)}+\chi_{v,1},\\
0,&\text{otherwise}.
\end{cases}
\end{eqnarray}
Then, it is easy to see that the quantum representation of Eq.  \eqref{PoKK1} is isomorphic to Eq.\eqref{noncom1}.
By following this analysis, similar results can be achieved for  the Poisson brackets $\{{e^{\pm\textbf{i}r(\eta_{e})\Xi^{\pm}_{e}(\vec{\chi})\cdot\varrho_{e}}},{e^{\pm\textbf{i}r(\eta_{e'})\Xi^{\pm}_{e'}(\vec{\chi})\cdot\varrho_{e'}}}\} $, $\{{e^{\pm\textbf{i}r(\eta_{v,I})\Xi^{\pm}_{v,I}(\vec{\chi})\cdot\varrho_{v,I}}},{e^{\pm\textbf{i}r(\eta_{e'})\Xi^{\pm}_{e'}(\vec{\chi})\cdot\varrho_{e'}}}\} $ and  $\{{e^{\pm\textbf{i}r(\eta_{v,I})\Xi^{\pm}_{v,I}(\vec{\chi})\cdot\varrho_{v,I}}},{e^{\pm\textbf{i}r(\eta_{v',I'})\Xi^{\pm}_{v',I'}(\vec{\chi})\cdot\varrho_{v',I'}}}\} $ .

Now, we can conclude that the quantum algebras among the operators  $\widehat{e^{\pm\textbf{i}r(\eta_e)\varrho_e}},\hat{\chi}_{e'},\widehat{e^{\pm\textbf{i}r(\eta_e){\varrho}_{v,I}}},\hat{\chi}_{v',I'}$ give a faithful quantum representation of the Poisson algebra among ${e^{\pm\textbf{i}r(\eta_e)\Xi^{\pm}_{e}(\vec{\chi})\varrho_e}}, {\chi}_{e'}, {e^{\pm\textbf{i}r(\eta_{v,I})\Xi^{\pm}_{v,I}(\vec{\chi}){\varrho}_{v,I}}}$ and ${\chi}_{v',I'}$ for some  specific regulators $\Xi^{\pm}_{e}(\vec{\chi})$ and $\Xi^{\pm}_{v,I}(\vec{\chi})$. By following the Eqs.\eqref{reg1}, \eqref{Poxixi} and \eqref{PoKK1}, one can choose the regulators $\Xi^{\pm}_{e}(\vec{\chi})$ and $\Xi^{\pm}_{v,I}(\vec{\chi})$  ensuring
\begin{equation}
1-t^n\leq \Xi^{\pm}_{e}(\vec{\chi}),\Xi^{\pm}_{v,I}(\vec{\chi})\leq1+t^n,
\end{equation}
with $t=\frac{\kappa\hbar}{a^2}$ and $n\gg1$ being a finite large real number. Then, one has
\begin{equation}
\frac{(e^{\textbf{i}r(\eta_e)\varrho_e}-e^{-\textbf{i}r(\eta_e)\varrho_e})}{2\textbf{i}}=\frac{(e^{\textbf{i}r(\eta_e) \Xi^{+}_{e}(\vec{\chi})\varrho_e}-e^{-\textbf{i}r(\eta_e) \Xi^{-}_{e}(\vec{\chi})\varrho_e})}{2\textbf{i}}+\mathcal{O}(t^n)
\end{equation}
and 
\begin{equation}
\frac{(e^{\textbf{i}r(\eta_{v,I})\varrho_{v,I}}-e^{-\textbf{i}r(\eta_{v,I})\varrho_{v,I}})}{2\textbf{i}}=\frac{(e^{\textbf{i}r(\eta_{v,I}) \Xi^{+}_{v,I}(\vec{\chi})\varrho_{v,I}}-e^{-\textbf{i}r(\eta_{v,I}) \Xi^{-}_{v,I}(\vec{\chi})\varrho_{v,I}})}{2\textbf{i}}+\mathcal{O}(t^n).
\end{equation}
Hence, though the operators  $\widehat{e^{\pm\textbf{i}r(\eta_e)\varrho_e}},\hat{\chi}_{e'},\widehat{e^{\pm\textbf{i}r(\eta_e){\varrho}_{v,I}}},\hat{\chi}_{v',I'}$ give a faithful quantum representation of ${e^{\pm\textbf{i}r(\eta_e)\Xi^{\pm}_{e}(\vec{\chi})\varrho_e}}, {\chi}_{e'}, {e^{\pm\textbf{i}r(\eta_{v,I})\Xi^{\pm}_{v,I}(\vec{\chi}){\varrho}_{v,I}}}$ and ${\chi}_{v',I'}$, they can also be regarded as the quantum representation of ${e^{\pm\textbf{i}r(\eta_e)\varrho_e}}, {\chi}_{e'}, {e^{\pm\textbf{i}r(\eta_{v,I}){\varrho}_{v,I}}}$ and ${\chi}_{v',I'}$ up to $\mathcal{O}(t^n)$.

\end{document}